\documentclass[11pt,a4paper]{article}
\pdfoutput=1

\usepackage{jheppub}
\usepackage{latexsym}
\usepackage{revsymb}
\usepackage{multirow}
\usepackage{color}
\usepackage[usenames,dvipsnames,svgnames,table]{xcolor}

\usepackage{graphicx}
\usepackage{epsfig}  
\usepackage{epsf}    
\usepackage{dcolumn}
\usepackage{bm}
\usepackage{dcolumn}
\usepackage{textcomp}
\usepackage{float}
\usepackage{subfig}
\usepackage{hypcap}
\usepackage[]{hyperref}

\hypersetup{
  bookmarks=true,         
  unicode=false,          
  pdftoolbar=true,        
 pdfmenubar=true,        
 pdffitwindow=true,     
 pdfstartview={FitH},    
 pdfsubject={Neutrino Oscillations Phenomenology},   
 pdfnewwindow=true,      
 pdfcreator={RevTeX},
 colorlinks=true,       
 linkcolor=red,          
 citecolor=blue,        
filecolor=black,      
 urlcolor=blue,           
  }

\def\nue{{\nu_e}}

\def\numu{{\nu_{\mu}}}

\def\nutau{{\nu_{\tau}}}

\def\anu{{\bar\nu}}

\newcommand{\beq}{\begin{equation}}
\newcommand{\eeq}{\end{equation}}
\newcommand{\beqa}{\begin{eqnarray}}
\newcommand{\eeqa}{\end{eqnarray}}

\newcommand{\tx}{{\theta_{12}}}
\newcommand{\ty}{{\theta_{13}}}
\newcommand{\tz}{{\theta_{23}}}
\newcommand{\sch}{\sin^2 \theta_{13}}
\newcommand{\stch}{\sin^2 2\theta_{13}}
\newcommand{\sa}{\sin^2 \theta_{23}}
\newcommand{\sta}{\sin^22 \theta_{23}}

\newcommand{\dl}{{\Delta_{31}}}
\newcommand{\ds}{{\Delta_{21}}}

\newcommand{\dcp}{\delta_{\mathrm{CP}}}
\newcommand{\nova}{NO$\nu$A~}
\newcommand{\pmue}{P(\nu_\mu \rightarrow \nu_e)}

\newcommand{\dchsq}{\Delta\chi^2}

\newcommand{\ie}{{\it i.e.}}

\preprint{IFIC-12-85}

\title{Resolving the octant of $\theta_{23}$ with T2K and NO$\nu$A}

\author[a,b]{Sanjib Kumar Agarwalla,}
\author[c]{Suprabh Prakash,} 
{\author[c]{S. Uma Sankar$\,$}

\affiliation[a]{Institute of Physics, Sachivalaya Marg, Sainik School Post, Bhubaneswar 751005, India}
\affiliation[b]{Instituto de F\'{\i}sica Corpuscular, CSIC-Universitat de Val\`encia, \\
         Apartado de Correos 22085, E-46071 Valencia, Spain}
\affiliation[c]{Department of Physics, Indian Institute of Technology Bombay, Mumbai 400076, India}

\emailAdd{sanjib@iopb.res.in}
\emailAdd{suprabh@phy.iitb.ac.in}
\emailAdd{uma@phy.iitb.ac.in}

\abstract
{
Preliminary results of MINOS experiment indicate that $\tz$ is not maximal. Global fits to world
neutrino data suggest two nearly degenerate solutions for $\tz$: one in the lower octant (LO: $\tz<45^\circ$)
and the other in the higher octant (HO: $\tz>45^\circ$). $\nu_{\mu}\rightarrow\nu_{e}$ oscillations in 
superbeam experiments are sensitive to the octant and are capable of resolving this degeneracy.
We study the prospects of this resolution by the current T2K and upcoming NO$\nu$A experiments.
Because of the hierarchy-$\dcp$ degeneracy and the octant-$\dcp$ degeneracy, the impact of
hierarchy on octant resolution has to be taken into account.
As in the case of hierarchy determination, there exist favorable (unfavorable) values of $\dcp$
for which octant resolution is easy (challenging). However, for octant resolution
the unfavorable $\dcp$ values of the neutrino data are favorable for the anti-neutrino data and vice-verse.
This is in contrast to the case of hierarchy determination.
In this paper, we compute the combined sensitivity of T2K and NO$\nu$A to resolve the octant ambiguity.
If $\sin^2\tz=0.41$, then NO$\nu$A can rule out all the values of $\tz$ in HO at $2\sigma$ C.L., 
irrespective of the hierarchy and $\dcp$. 
Addition of T2K data improves the octant sensitivity. 
If T2K were to have equal neutrino and anti-neutrino runs of 2.5 years each,
a $2\sigma$ resolution of the octant becomes possible provided $\sin^2\tz\leq0.43~\textrm{or}~ \geq0.58$
for any value of $\dcp$.
}


\begin{document}
\maketitle
\flushbottom

\section{Introduction and Motivation}
\label{introduction}

Our understanding of the smallest lepton mixing angle $\ty$ has improved quite dramatically in last one year or so and
finally it has been confirmed to be non-zero with unprecedented confidence by the reactor experiments 
Daya Bay \cite{An:2012eh} and RENO \cite{Ahn:2012nd}.
They have found a reasonably large 1-3 mixing
\begin{center}
\hskip0.5cm$\stch|_{\mathrm{\bf Daya Bay}}$ = $0.089 \pm 0.010 \,
({\mathrm{stat}}) \pm 0.005 \, ({\mathrm{syst}})$ \cite{Kyoto2012DayaBay}, and
\vskip0.2cm$\left.\stch\right|_{\mathrm{\bf RENO}}$ = $0.113 \pm 0.013 \, ({\mathrm{stat}}) \pm 0.019 \, ({\mathrm{syst}})$ \cite{Kyoto2012RENO},
\end{center}
in agreement with the measurements performed earlier by T2K \cite{Abe:2011sj,Kyoto2012T2K}, MINOS \cite{Adamson:2011qu,Nichol:2012},
and Double Chooz \cite{Abe:2011fz,Abe:2012tg} experiments.
Combined analyses of all the neutrino oscillation data available \cite{Tortola:2012te,Fogli:2012ua,GonzalezGarcia:2012sz} imply a non-zero value of $\ty$ 
at more than $10\sigma$ and predict a best-fit value of $\sch \simeq 0.023$ with a relative $1\sigma$ precision of 10\%. 
These recent high precision measurements of $\ty$ have taken us one step further in validating the standard three flavor oscillation picture of 
neutrinos on a strong footing \cite{Hewett:2012ns}. Also, a moderately large value of $\ty$ has provided a {\it `golden'} opportunity 
to directly determine the neutrino mass hierarchy\footnote{Two possibilities are there: it can be either 
normal (NH) if $\Delta_{31} \equiv m^2_3 -m^2_1 > 0$, or inverted (IH) if $\Delta_{31} < 0$.} (NMH) using the Earth matter effects, and to unravel 
the evidence of leptonic CP violation (LCPV)\footnote{If the Dirac CP phase, $\dcp$ differs from 0 or $180^\circ$.} in accelerator based 
long-baseline neutrino oscillation experiments \cite{Kyoto2012Minakata}.

Another recent and crucial development is the indication of non-maximal 
$\tz$ by the MINOS accelerator experiment \cite{Nichol:2012}. However, the atmospheric neutrino data, dominated by 
Super-Kamiokande, still prefers the maximal value of $\tz$ \cite{Itow:2012}. All the three global fits of 
world neutrino data \cite{Tortola:2012te,Fogli:2012ua,GonzalezGarcia:2012sz} also point to the 
deviation from maximal mixing for $\tz$ \ie\,\,$(0.5 - \sa) \ne 0$.

Both these new measurements, non-zero value of $\ty$ and non-maximal $\tz$, will provide crucial inputs to the theories 
of neutrino masses and mixings \cite{Mohapatra:2006gs,Albright:2006cw,Albright:2010ap,King:2013eh}. 
A number of innovative ideas, such as $\mu \leftrightarrow \tau$ symmetry 
\cite{Fukuyama:1997ky,Mohapatra:1998ka,Lam:2001fb,Harrison:2002et,Kitabayashi:2002jd,Grimus:2003kq,Ghosal:2003mq,Koide:2003rx,Mohapatra:2005yu}, 
$A_4$ flavor symmetry \cite{Ma:2002ge,Ma:2001dn,Babu:2002dz,Grimus:2005mu,Ma:2005mw},
and quark-lepton complementarity \cite{Raidal:2004iw,Minakata:2004xt,Ferrandis:2004vp,Antusch:2005ca} have been invoked to explain the 
observed pattern of one small and two large mixing angles in the neutrino sector. Measurements of the precise values of $\ty$ and $\tz$ 
will reveal the pattern of deviations from these symmetries and will lead to a better understanding of neutrino masses and mixings. 
In particular, the resolution of $\tz$ octant will severely constrain the patterns of symmetry breaking.

In $\numu$ survival probability, the dominant term is mainly sensitive to $\sta$. 
Now, if $\sta$ differs from 1 as indicated by the recent neutrino data, then we get two solutions for $\tz$: 
one $< 45^\circ$, termed as lower octant (LO) and the other $> 45^\circ$, termed as higher octant (HO). 
In other words, if the quantity $(0.5 - \sa)$ is positive (negative) then $\tz$ belongs to LO (HO).
This is known as the octant degeneracy of $\tz$ \cite{Fogli:1996pv} which is a part of the overall eight-fold degeneracy \cite{Barger:2001yr,Minakata:2002qi}, 
where the other two degeneracies are $(\ty,\dcp)$ intrinsic degeneracy \cite{BurguetCastell:2001ez} and the 
(hierarchy, $\dcp$) degeneracy \cite{Minakata:2001qm}.

The octant ambiguity of $\tz$ is considered to be the most difficult one to deal with among the eight-fold parameter degeneracies. 
In the past when we had only an upper bound on $\ty$, a possible way of resolving this degeneracy by combining future reactor data with 
accelerator $\numu$ disappearance and $\nue$ appearance measurements was suggested in \cite{Minakata:2002jv,Hiraide:2006vh}.
Adding the information from the `silver' channel $(\nue \to \nutau)$ to the `golden' channel $(\nue \to \numu)$ in the proposed neutrino
factory setup is demonstrated to be one of the elegant ways to tackle this degeneracy \cite{Donini:2002rm,Meloni:2008bd}. 
The possibility of determining the deviation of $\theta_{23}$ from maximal mixing and consequently the correct octant of $\theta_{23}$ 
in very long-baseline neutrino oscillation experiments and as well as in future atmospheric neutrino experiments has been discussed in 
\cite{Antusch:2004yx,Minakata:2004pg,GonzalezGarcia:2004cu,Choudhury:2004sv,Choubey:2005zy,Indumathi:2006gr,Kajita:2006bt,Hagiwara:2006nn,Samanta:2010xm}.  
One clear message that has been conveyed by all these novel works is that one can achieve a very good sensitivity to the quantity 
$|0.5 - \sa|$ from the conventional beam experiments (MINOS, ICARUS and OPERA), the current generation superbeam experiments 
(presently running T2K and upcoming NO$\nu$A), and also from the current (Super-Kamiokande) and future atmospheric 
data (India-based Neutrino Observatory). But, determining the sign of $|0.5 - \sa|$ is deemed to be a very difficult job to pursue and 
it demands a large value of $\ty$. 

Now, in the light of recently discovered moderately large value of $\ty$, it would be quite interesting to study whether the expected appearance 
data from the ongoing T2K experiment \cite{Itow:2001ee,Kyoto2012T2K} in Japan and the upcoming NO$\nu$A 
experiment \cite{Ayres:2004js,Kyoto2012nova} in the United States can resolve the octant ambiguity of $\tz$ or not?
In this paper, we address this issue.

The structure of the paper is as follows. We start in section \ref{currentstatus} by revisiting our present
understanding of the 2-3 mixing angle. Section \ref{physics} describes in detail the physics issues related 
to the octant of $\tz$. We show the event rates for T2K and NO$\nu$A in section \ref{events}. At the end of this
section, we also describe the simulation method followed. 
We present our results in section \ref{results}. Finally, in section \ref{summary}, we summarize and draw our conclusions.
Expected events rates in T2K and NO$\nu$A (both for neutrino and anti-neutrino) as a function of $\dcp$ can be found in
Appendix \ref{appendix1}. Allowed regions in the $\sa$(test) - $\dcp$(test) plane for the true value of $\dcp = 0$
and all the four combinations of true hierarchy and true octant are shown in Appendix \ref{errorplots}.

\section{Present Understanding of the 2-3 mixing angle}
\label{currentstatus}

Our present knowledge of $\tz$ comes from two sources: a) atmospheric neutrinos
and b) accelerator neutrinos. In both cases, the muon neutrino disappearance is 
parametrized in the form of two-flavor survival probability 
\beq
\textrm{P}_{\mu \mu}= 1-\sin^22\theta_{\textrm{eff}}\sin^2\left(\frac{\Delta m^2_{\textrm{eff}}L}{4E}\right).
\eeq
Analysis of the data gives reasonably precise values for the effective two-flavor parameters.
Relating these to the three flavor parameters depends on the experimental set up.
In the case of atmospheric neutrinos, the path lengths involved vary from 20 km to 
13000 km and the energies vary from 200 MeV to a few GeV. This represents a 
very wide range in L/E. The approximations valid 
for some values of L/E are not valid for others. Therefore, for atmospheric neutrinos,
it is not possible to obtain a direct relation
between the effective two-flavor parameters and the three-flavor parameters.
However, for accelerator neutrinos, L and E are chosen so that $\Delta_{31}L/E\sim 90^\circ$.
Hence, $\Delta_{21}L/E \ll 1$ ($\Delta_{21}=m^2_2-m^2_1$) and can be treated as a small perturbation. In this approximation, 
it was shown that \cite{Nunokawa:2005nx,deGouvea:2005hk,Raut:2012dm}
\beqa
\Delta m^2_{\textrm{eff}}&=& \dl - \left(\cos^2\tx - \cos \dcp \sin\ty \sin 2\tx \tan \tz\right)\ds, \label{d31eff} \\
\sin^22\theta_{\textrm{eff}}&=& 4\cos^2\ty\sin^2\tz\left(1-\cos^2\ty\sin^2\tz\right) \label{tzeff}.
\eeqa
The mixing angles and phases are defined according to the Particle Data Group convention \cite{Beringer:1900zz}.
The atmospheric neutrino data, dominated by Super-Kamiokande, still prefers the maximal value of 
$\sin^22\theta_{\textrm{eff}}=1\,(\ge 0.94\,(90\%\,\mathrm{C.L.}))$ \cite{Itow:2012}. 
But the preliminary results from the MINOS accelerator experiment favor a non-maximal value of    
$\sin^2 2\theta_{\textrm{eff}}=0.94^{+0.04}_{-0.05}$ \cite{Nichol:2012}.

\begin{table}[H]
\begin{center}
\scalebox{0.9}{
\begin{tabular}{|l||c|c|c|}
\hline
Reference
& Forero et.al.~\cite{Tortola:2012te}
& Fogli et.al.~\cite{Fogli:2012ua}
& Gonzalez-Garcia et.al.~\cite{GonzalezGarcia:2012sz}
\\
\hline\hline
$\sin^2\theta_{23}$ (NH)
& $0.427^{+0.034}_{-0.027}\oplus 0.613^{+0.022}_{-0.040}$
& $0.386^{+0.024}_{-0.021}$
& $0.41^{+0.037}_{-0.025}\oplus 0.59^{+0.021}_{-0.022}$
\\
$3\sigma$ range
& $0.36\rightarrow 0.68$
& $0.331\rightarrow 0.637$
& $0.34\rightarrow 0.67$
\\
\cline{1-3}
$\sin^2\theta_{23}$ (IH)
& $0.600^{+0.026}_{-0.031}$
& $0.392^{+0.039}_{-0.022}$
&
\\
$3\sigma$ range
& $0.37\rightarrow 0.67$
& $0.335\rightarrow 0.663$
&
\\
\hline\hline
\end{tabular}
}
\caption{{\footnotesize $1\sigma$ bounds on $\sin^2\theta_{23}$ from the global fits performed in
References~\cite{Tortola:2012te}, \cite{Fogli:2012ua}, and \cite{GonzalezGarcia:2012sz}.
NH and IH stand for normal and inverted hierarchies. The numbers cited from Ref.~\cite{GonzalezGarcia:2012sz} 
are those obtained by keeping the reactor fluxes free in the fit and also including the short-baseline reactor
data with $L \lesssim 100$ m, with the mass hierarchy marginalized.}}
\label{sin2thetaXY}
\end{center}
\end{table}

Global fits, using three-flavor oscillations, give information directly on $\tz$ rather than $\theta_{\textrm{\textrm{eff}}}$.
The best-fit values and ranges of $\tz$ from the three recent global fits \cite{Tortola:2012te}, \cite{Fogli:2012ua}, and 
\cite{GonzalezGarcia:2012sz} are listed in Table \ref{sin2thetaXY}.
A common feature that has emerged from all the three global fits of the world neutrino data is that we now have indication
for non-maximal $\theta_{23}$. Thus, we have the two degenerate solutions: either $\tz$ belongs to the LO
($\sin^2\theta_{23} \approx 0.4$) or it lies in the HO ($\sin^2\theta_{23}\approx0.6$).
This degeneracy, in principle, can be broken with the help of $\nu_{\mu} \leftrightarrow \nu_{e}$ oscillation data.
The preferred value would depend on the choice of the neutrino mass hierarchy.
However, as can be seen from Table \ref{sin2thetaXY}, the fits of reference \cite{Tortola:2012te} do not agree on which value 
should be preferred, even when the mass hierarchy is fixed to be NH.
In \cite{Fogli:2012ua}, LO is preferred over HO for both NH and IH. 
Reference \cite{GonzalezGarcia:2012sz} marginalizes over the mass hierarchy, so the degeneracy remains. 
The global best-fits in references \cite{Tortola:2012te,GonzalezGarcia:2012sz} do not observe any sensitivity to the
octant of $\tz$ unless they add the atmospheric neutrino data. But, in \cite{Fogli:2012ua}, they do find a preference
for LO even without adding the atmospheric data. In this paper, we take the best-fit value of $\sin^2\tz$ in the
lower octant (LO) to be 0.41 while that in the higher octant (HO) to be 0.59 \cite{GonzalezGarcia:2012sz}.
 
\section{Physics of the octant of $\tz$}
\label{physics}

In a long-baseline experiment, $\nu_{\mu}$ charged current (CC) events are the most copious.
These experiments can measure $\nu_{\mu}\rightarrow\nu_{\mu}$ survival probability, $\textrm{P}_{\mu\mu}$, as a
function of energy. The reconstruction of the minimum of $\textrm{P}_{\mu\mu}$ leads to precise 
values for $|\Delta m^2_{\textrm{eff}}|$ and $\sin^22\theta_{\textrm{eff}}$ \cite{Nunokawa:2005nx,deGouvea:2005hk,Raut:2012dm,Nichol:2012}.
Therefore, we get two degenerate best-fit values for $\sin^2\tz$, one in LO and the other in HO, with 
small allowed regions around them. The $\textrm{P}_{\mu\mu}$ expression has subleading terms which are 
sensitive to octant \cite{Akhmedov:2004ny}. But, these are suppressed by the small parameter 
$\alpha = \ds/\dl$ and their biggest impact occurs
for energies where $\textrm{P}_{\mu\mu}$ is very small. So, the overall octant sensitivity of $\textrm{P}_{\mu\mu}$
is negligible. However, $\textrm{P}_{\mu\mu}$ gives a precise measurement of $\sin^22\theta_{\textrm{eff}}$ which
in turn gives precise allowed regions for $\sin^2\tz$ in the two octants.

In the presence of matter, the $\nu_{\mu} \to \nu_{e}$ oscillation 
probability, expanded perturbatively in $\alpha$ and $\ty$,  
can be written as \cite{Cervera:2000kp,Akhmedov:2004ny,Freund:2001pn} 
\beq
P_{\mu e}= \beta_{1}\sin^2\tz + \beta_{2}\cos(\hat{\Delta}+\dcp) + \beta_{3}\cos^2\tz.
\label{pmue23}
\eeq
In the above equation, $P_{\mu e}$ is written in a way to highlight the octant and $\dcp$ dependent terms. Here
\beqa
\beta_{1} &=& \sin^2 2\ty
\frac{\sin^2\hat{\Delta}(1 - \hat{A})}{(1 - \hat{A})^2},  \nonumber \\ 
\beta_{2} &=& \alpha\cos\ty\sin2\tx\sin2\ty\sin2\tz\frac{\sin\hat{\Delta}\hat{A}}{\hat{A}}
\frac{\sin\hat{\Delta}(1-\hat{A})}{1-\hat{A}}, \nonumber \\
\beta_{3} &=& \alpha^2\sin^22\tx\cos^2\ty\frac{\sin^2\hat{\Delta}\hat{A}}{\hat{A}^2},
\eeqa
with $\hat{\Delta}= \dl L/4E$, $\hat{A}= A/\dl$.
$A$ is the Wolfenstein matter term \cite{msw1} and is given by
$A ({\rm eV}^2) = 0.76 \times 10^{-4} \rho \ ({\rm g/cc}) E ({\rm GeV})$.
$\rho$ is the density of matter in the Earth. For \nova and T2K, this is 
set equal to the density in the crust of 2.8 g/cc.

\begin{figure}[H]

        \begin{tabular}{lr}
                \hspace*{-0.85in} \includegraphics[width=0.75\textwidth]{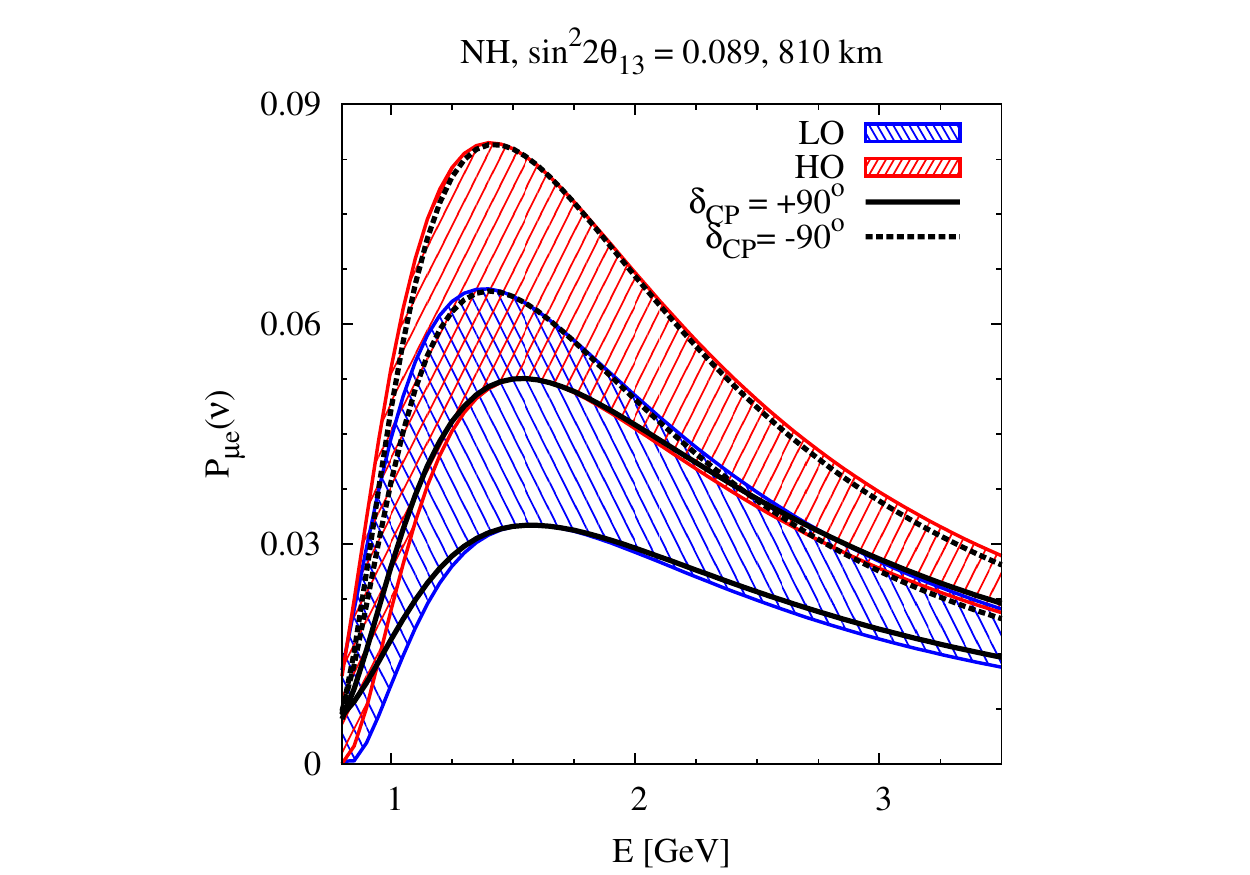}
                & 
                \hspace*{-1.7in} \includegraphics[width=0.75\textwidth]{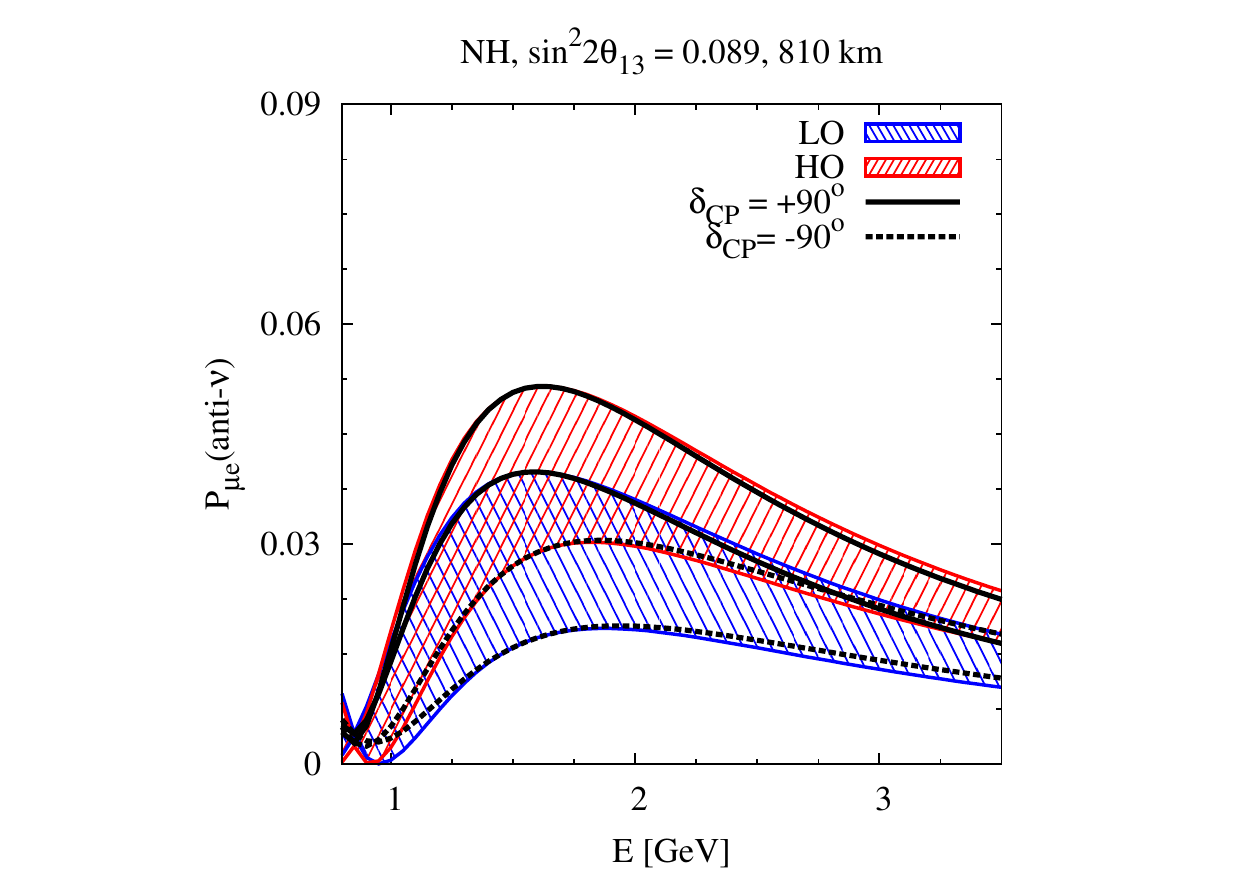}
        \end{tabular}
        
\caption{\label{pmue_octant_dcp_band_NH}\footnotesize $P_{\mu e}$ as a function of neutrino energy.
The left panel (right panel) is for $\nu$ ($\anu$). Here, the bands correspond to different 
values of $\dcp$ from $-180^\circ$ to $180^\circ$. These plots are for \nova (L=810 km), $\sin^22\ty = 0.089$ 
and NH. For LO (HO), $\sin^2\tz = 0.41~(0.59)$.}

\end{figure}

For normal hierarchy (NH), $\dl$ is positive and for inverted hierarchy (IH),
it is negative. The matter term $A$ is positive for neutrinos
and is negative for anti-neutrinos. Hence, in neutrino oscillation
probability, $\hat{A}$ is positive for NH and is negative for IH; vice-verse for anti-neutrinos.
Moreover, sign of $\dcp$ is reversed for anti-neutrinos. 
The left (right) panel of figure \ref{pmue_octant_dcp_band_NH} shows
$P_{\mu e}$ vs. $\textrm{E}_{\nu}$ ($P_{\bar{\mu}\bar{e}}$ vs. $\textrm{E}_{\bar{\nu}}$)
for all possible values of $\dcp$ and for the two values of $\sin^2\tz$, assuming
NH to be the true hierarchy. These plots are for the experiment \nova.

As can be seen from the left panel of figure \ref{pmue_octant_dcp_band_NH}, for neutrino data, 
the two octant bands overlap for some values of $\dcp$ and are
distinct for other values. The combinations of octant and $\dcp$
which lie farthest from overlap will be favorable combinations
for octant determination. For example, LO and $\dcp$ of $90^\circ$ 
and HO and $\dcp$ of $-90^\circ$ form the favorable combinations. For
the combinations with overlap, HO and $\dcp$ of $90^\circ$ and LO
and $\dcp$ of $-90^\circ$, it is impossible to determine octant using
neutrino data alone. However, as we see from the right panel, these
unfavorable combinations for neutrino case are the favorable ones for the
$\bar{\nu}$ case. Thus, a combination of neutrino and anti-neutrino data
will have a better capability to determine octant compared to neutrino data
alone. This is in contrast to the hierarchy-$\dcp$ degeneracy, where for a given hierarchy, 
the favorable $\dcp$ region is the same for both $\nu$ and $\anu$.
Thus, we are led to the conclusion that a balanced neutrino
and anti-neutrino data is imperative for resolving the octant for all values of $\dcp$.

Let us do a small, quantitative analysis of the octant-$\dcp$ degeneracy of $\textrm{P}_{\mu e}$.
A similar analysis for hierarchy-$\dcp$ degeneracy was done in \cite{Mena:2004sa}.
For simplicity, here we keep the hierarchy fixed.
Increase in $\tz$ increases $P_{\mu e}$. While a change in $\dcp$ can increase or decrease $P_{\mu e}$.
For different $\dcp^{\textrm{LO}}$ and $\dcp^{\textrm{HO}}$, 
$P_{\mu e}(\textrm{LO}, \dcp^{\textrm{LO}})$ may be very close to $P_{\mu e}(\textrm{HO}, \dcp^{\textrm{HO}})$.
For the degenerate case, $P_{\mu e}(\textrm{LO}, \dcp^{\textrm{LO}})=P_{\mu e}(\textrm{HO}, \dcp^{\textrm{HO}})$,
leading to
\beq
\cos(\hat{\Delta}+\dcp^{\textrm{LO}}) - \cos(\hat{\Delta}+\dcp^{\textrm{HO}}) = 
\frac{\beta_1 - \beta_3}{\beta_2}(\sin^2\theta_{23}^{\textrm{HO}} - \sin^2\theta_{23}^{\textrm{LO}}).
\eeq
For the \nova baseline L = 810 km and energy of peak flux E = 2 GeV, we get for NH and $\nu$
\beq
\cos(\hat{\Delta}+\dcp^{\textrm{LO}}) - \cos(\hat{\Delta}+\dcp^{\textrm{HO}}) = 1.7.
\label{NHnudegenerate}
\eeq
The above equation will have solutions only if 
\beqa
0.7&\leq& \cos(\hat{\Delta}+\dcp^{\textrm{LO}}) \leq 1.0, \nonumber \\
-1.0&\leq& \cos(\hat{\Delta}+\dcp^{\textrm{HO}}) \leq -0.7. 
\eeqa
From this, we get their ranges to be:
\beqa
-116^\circ \leq &\dcp^{\textrm{LO}}& \leq -26^\circ, \nonumber \\
64^\circ \leq &\dcp^{\textrm{HO}}& \leq 161^\circ. 
\eeqa
Thus, we find that for NH and $\nu$ of energy 2 GeV, 
$P_{\mu e}(\textrm{LO}, -116^\circ \leq \dcp \leq -26^\circ)$  is degenerate with 
$P_{\mu e}(\textrm{HO}, 64^\circ \leq \dcp \leq 161^\circ)$.
A similar analysis can be done for T2K baseline of L = 295 km and energy of peak flux E = 0.6 GeV.
The overlap regions in the $\dcp$ range are essentially the same as above because the values
of $\hat{\Delta}$ and those of $(\beta_1 - \beta_3)/\beta_2$ are nearly the same for the two experiments.
This is to be contrasted with the hierarchy discrimination, where the overlap range for \nova
is very different from that of T2K because of widely different matter effects \cite{Mena:2004sa,Prakash:2012az}.

\begin{figure}[H]

        \begin{tabular}{lr}
                \hspace*{-0.85in} \includegraphics[width=0.75\textwidth]{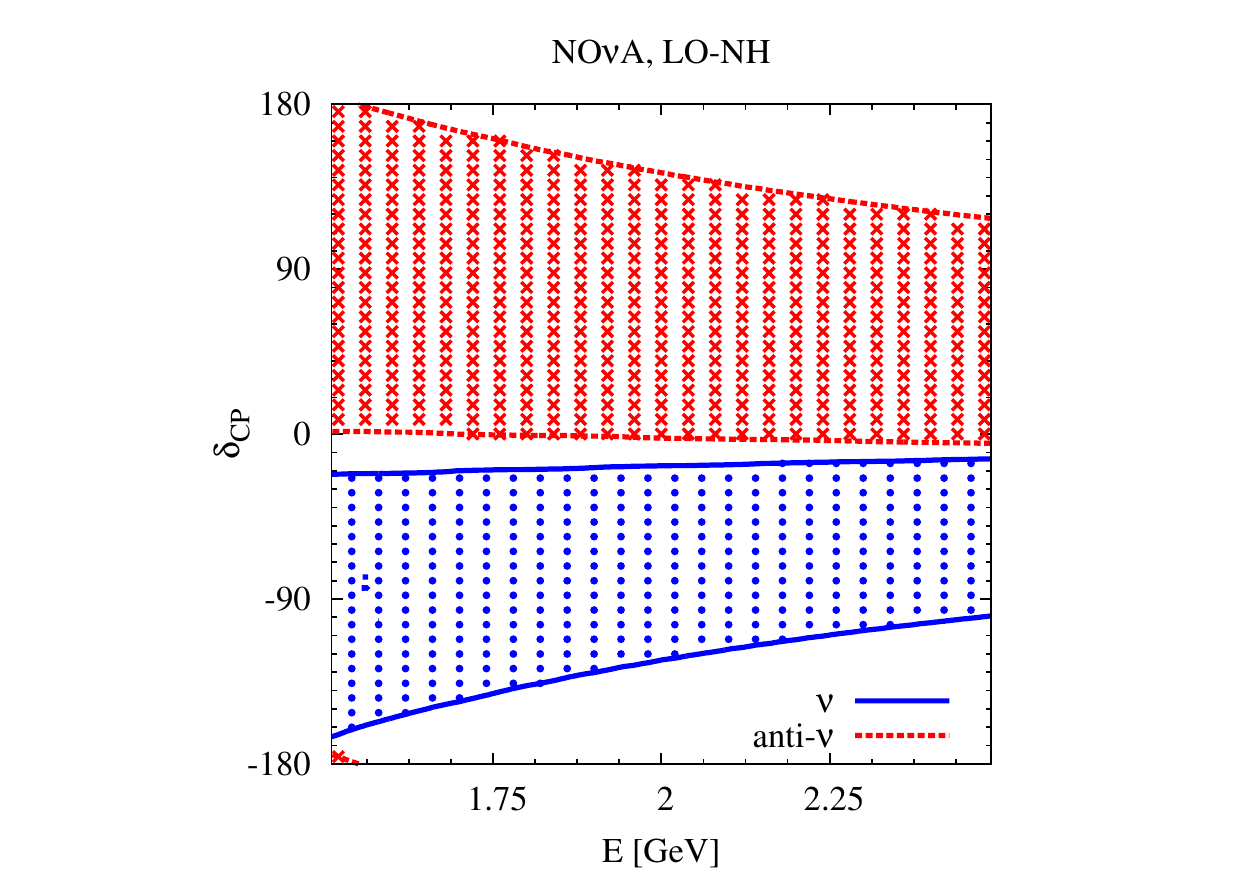}
                & 
                \hspace*{-1.7in} \includegraphics[width=0.75\textwidth]{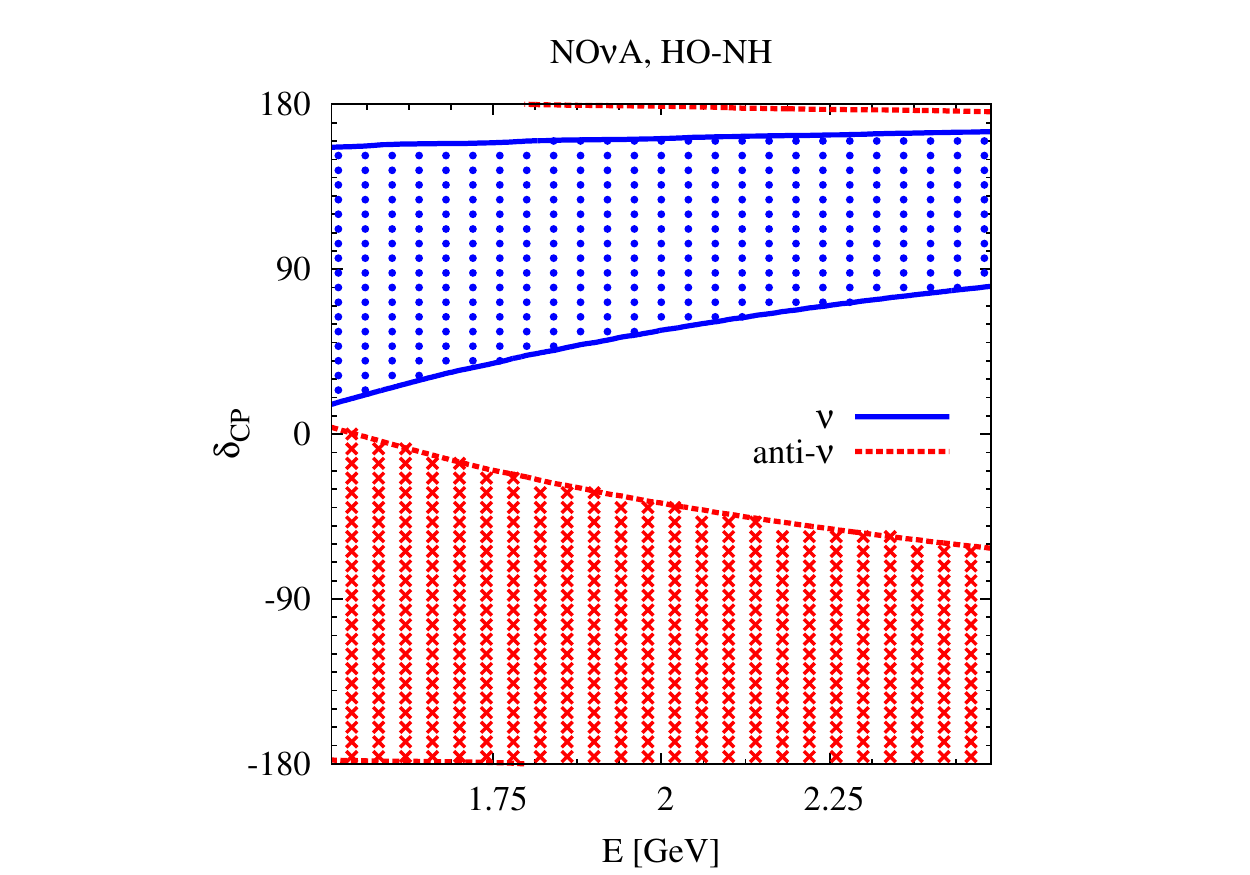}
        \end{tabular}
        
\caption{\label{pmue_overlap_NH}\footnotesize Plots showing octant-$\dcp$ degeneracy
in $P_{\mu e}$ as a function of neutrino energy. The left (right) panel is for
LO (HO). The blue-dotted (red-crossed) regions are for $\nu$ ($\anu$).
For a given $\textrm{E}_{\nu}$ ($\textrm{E}_{\anu}$), $P_{\mu e}$[ LO, {\it vertical blue-dotted (red-crossed) $\dcp$ region in
the left panel} ] values are degenerate with $P_{\mu e}$[ HO, {\it vertical blue-dotted (red-crossed) $\dcp$ region in the
right panel} ] values. The exact degenerate octant-$\dcp$ values can be found out using equation \ref{NHnudegenerate}. 
As an example, for $\textrm{E}_{\nu}$ of 2 GeV, $P_{\mu e}$(LO, $\dcp=-90^\circ$) is degenerate with 
$P_{\mu e}$(HO, $\dcp\approx66^\circ$). These plots are for \nova (L=810 km), 
$\sin^22\ty = 0.089$ and NH. For LO (HO), $\sin^2\tz = 0.41~(0.59)$. }

\end{figure}

In figure \ref{pmue_overlap_NH}, we show the values of 
$\dcp$, which lie in the overlapping region, as a function of energy, for the
combinations LO-NH and HO-NH. For the experiment \nova, most of the signal events come from the 
range 1.5 - 2.5 GeV.
Hence, we consider only this energy range in the figure.
The blue-dotted (red-crossed) shaded region shows those values of $\dcp$, for which 
there is an overlap between $P_{\mu e}^{\textrm{LO}}$ and $P_{\mu e}^{\textrm{HO}}$,
for $\nu$ ($\bar{\nu}$) data. These plots show the degenerate octant-$\dcp$ values in the
relevant energy range and help explain the $\dcp$ dependence seen in octant sensitivity.

The values of $\dcp$ which lie farthest
from any of the overlap regions will be the most favored ones. Those 
$\dcp$ values which lie in one of the overlap region (i.e. either in $\nu$
or $\anu$) but are far from the other overlap region will also be favored (though
to a lesser extent than the previous values) if balanced $\nu$ and $\anu$ runs are taken.
The regions of $\dcp$ which are common or very close to both overlap regions will 
be the most unfavored ones. We find that $\dcp \sim 0$ is the most unfavorable for LO-NH
and HO-IH whereas $\dcp \sim 180^\circ$ is the most unfavorable for HO-NH and
LO-IH. This pattern is observed in the results in subsection \ref{allowedtheta23plots}.

\section{Event Rates for T2K and NO$\nu$A}
\label{events}

In this paper, we simulate the data for the two {\it off-axis} superbeam experiments: T2K and \nova using GLoBES 
\cite{Huber:2004ka, Huber:2007ji}.
In the T2K experiment, a $\nu_{\mu}$ beam from J-PARC is directed towards Super-Kamiokande detector,
295 km away. The flux peaks sharply at 0.6 GeV, close to the first oscillation maximum in $\pmue$. 
The experiment is scheduled to run for 5 years in the neutrino mode only. 
The details of T2K experiment are given in \cite{Itow:2001ee}. The information
regarding signal efficiencies and backgrounds are taken from \cite{Huber:2009cw, fechnerthesis}.
\nova is a 14 kT totally active scintillator detector located at Ash River; a distance of 810 km from Fermilab. 
The flux peaks at 2 GeV, again close to the first oscillation maximum in $\pmue$. This experiment is
scheduled to have three years run in neutrino mode first and then later, three years run in anti-neutrino mode as well. 
The details of the experiment are given in \cite{nova_tdr}. In light of the recent measurement of large $\ty$, \nova 
has reoptimized their signal and background acceptances. In our calculations, we use these reoptimized values,
the details of which are given in \cite{Kyoto2012nova, Agarwalla:2012bv}.

In our simulations, we have used the following input values for neutrino oscillation parameters 
\cite{Nichol:2012,GonzalezGarcia:2012sz}.
\beqa
&& |\Delta m^2_{\mathrm{eff}}| = 2.4 \times 10^{-3}\,\mathrm{eV}^2 \nonumber \\
&& \ds = 7.5\times10^{-5}\,\mathrm{eV}^2, \hspace*{1cm} \sin^2 \theta_{12} = 0.3 \nonumber \\
&& \sin^22\ty = 0.089 \nonumber
\label{eq:param}
\eeqa

The value of $\dl$ is calculated 
separately for NH and for IH using equation \ref{d31eff} where $\Delta m^2_{\mathrm{eff}}$ is
taken to be +ve for NH and -ve for IH.
The uncertainties in the above parameters are taken to be
$\sigma(|\Delta m^2_{\mathrm{eff}}|) = 4\%$ \cite{Nichol:2012} and 
$\sigma(\sin^22\theta_{13}) = 5\%$ \cite{dayabay_NF12}.  
The solar parameters ($\ds$ and $\sin^2\tx$) and the Earth matter density in the calculation
of matter term are held fixed throughout the calculation. 

\begin{figure}[H]

        \begin{tabular}{lr}
                \hspace*{-0.85in} \includegraphics[width=0.75\textwidth]{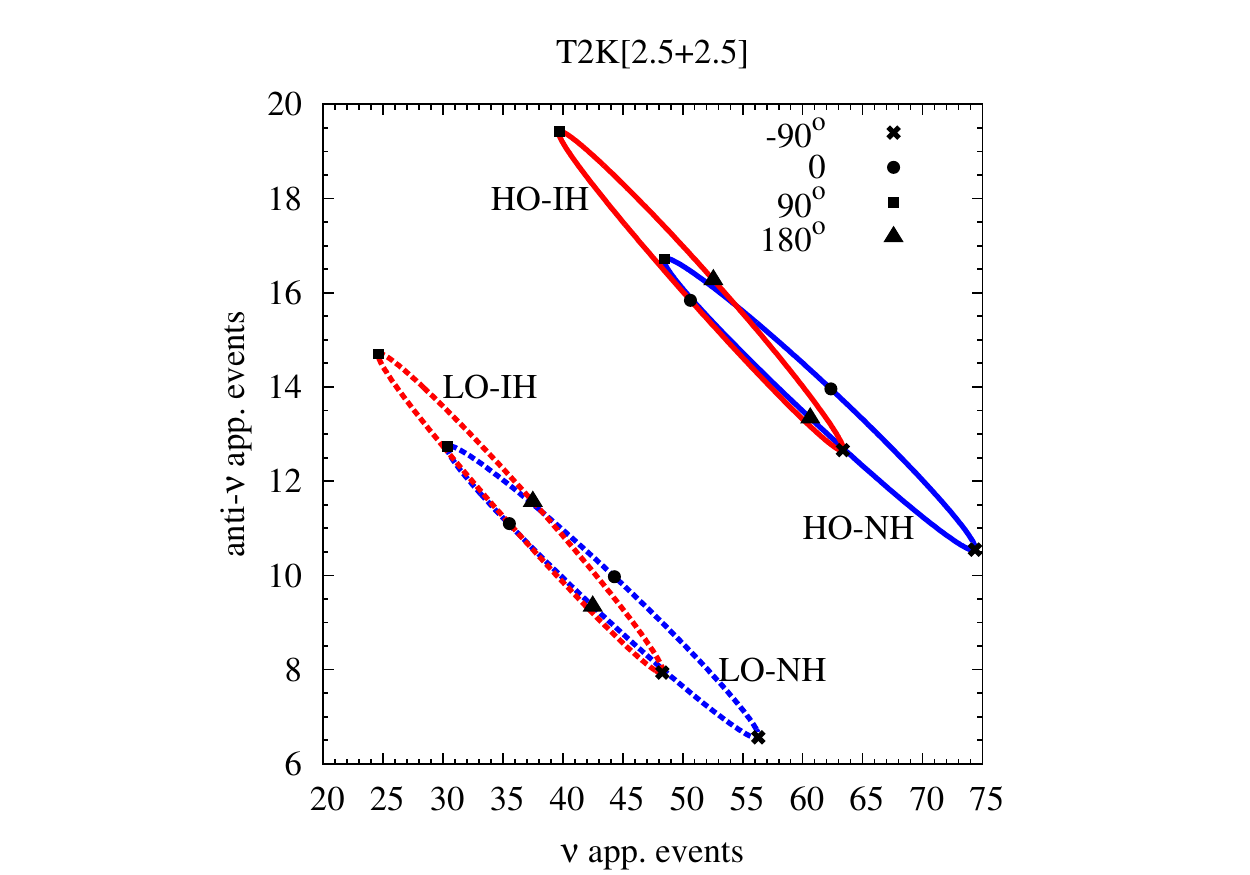}
                & 
                \hspace*{-1.7in} \includegraphics[width=0.75\textwidth]{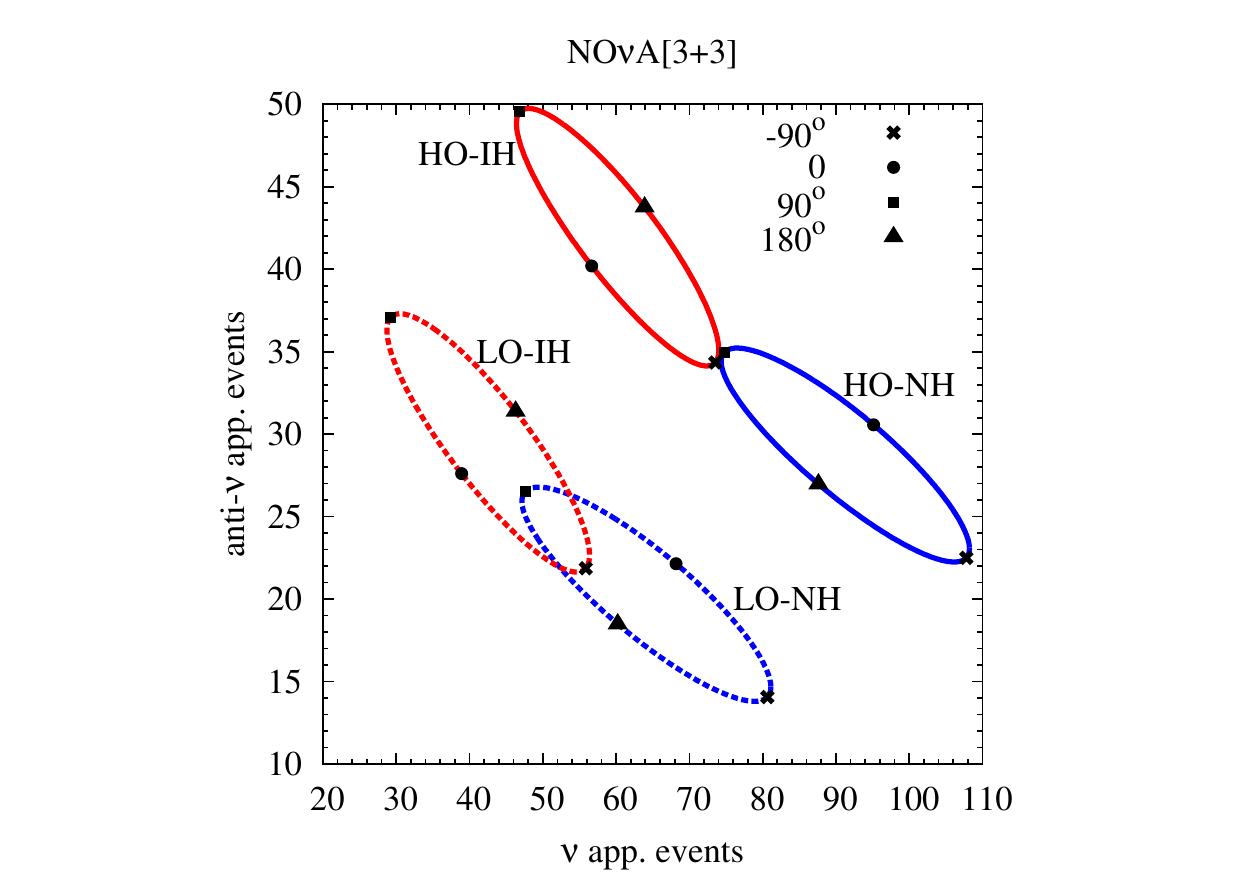}
        \end{tabular}
        
\caption{\label{iso-contour-event}\footnotesize $\nu$ and $\anu$ appearance events
for all possible combinations of hierarchy, octant and $\dcp$. The left (right) panel is for T2K 
(NO$\nu$A). Here $\sin^22\ty = 0.089$. For LO (HO), $\sin^2\tz = 0.41~(0.59)$. Note that for T2K,
equal $\nu$ and $\anu$ runs of 2.5 years each has been assumed. The variation of $\nu$ and $\anu$ appearance
events with $\dcp$ is shown in appendix \ref{appendix1} for both T2K and \nova.}

\end{figure}

In figure \ref{iso-contour-event},
we plot $\nu$ events vs. $\anu$ events
for various octant-hierarchy combinations. In each case, with varying values
of $\dcp$, the plot becomes an ellipse. The left panel shows these ellipses
for T2K whereas the right panel shows the same for \nova. {\it Here, we assumed that
T2K will have equal $\nu$ and $\anu$ runs of 2.5 years each}. In the right panel, 
we see that the ellipses for the two
hierarchies overlap whereas the ellipses of LO are well separated from those of
HO. Hence, we can expect that \nova will have better octant resolution capability
than hierarchy discrimination. This situation is even more dramatic in the left
panel where there is large overlap between the two hierarchies but clear separation
between the octants. Thus, it is very likely that $\anu$ data from T2K may play
an important role in the determination of octant.

We illustrate the octant determination capability of \nova by considering some special
points in the right panel. The features that we emphasize here can also be discerned from
figure \ref{pmue_overlap_NH}.

\begin{itemize}
 \item \underline{LO-NH, $\dcp=0^\circ$}: The coordinates of this point in the $\nu$-$\anu$ event plane are (68,22).
        No point on the HO-NH ellipse has these coordinates but there exist a number of points with 
        coordinates close to these. Therefore, the wrong octant is difficult to rule out for this point.
        
 \item \underline{LO-NH, $\dcp=90^\circ$}: The coordinates of this point in the $\nu$-$\anu$ event plane are (48,27).
       The $\nu$ events are much lower than those of all points on the HO-NH ellipse but there is degeneracy
       in the $\anu$ events. Hence, ruling out the wrong octant should be possible for this point.
       
 \item \underline{LO-NH, $\dcp=180^\circ$}: The coordinates of this point in the $\nu$-$\anu$ event plane are (60,18).
       Both $\nu$ and $\anu$ events are much lower than the corresponding events of any point on the HO-NH ellipse.
       Thus, ruling out the wrong octant will be the easiest for this point.
       
 \item \underline{LO-NH, $\dcp=-90^\circ$}: An argument similar to the case $\dcp=90^\circ$ can be made, except that the $\nu$ 
       events have degeneracy between octants, but the $\anu$ events for this point are much below those of any
       point on the HO-NH ellipse.

\end{itemize}

Similar arguments can be made for other octant-hierarchy combinations with the
exception that the most favorable and most unfavorable $\dcp$ values will differ.

We see that for the given choices of best-fit value of $\sin^2\tz$, \nova has very good sensitivity
to octant resolution due to balanced $\nu$ and $\anu$ runs. Because the favorable and 
unfavorable values of $\dcp$ (pertaining to octant resolution) are different for $\nu$ and $\anu$,
no $\dcp$ value is absolutely unfavorable if balanced $\nu$ and $\anu$ runs are taken.
Therefore, a single experiment on its own, can have very good sensitivity as illustrated in figure \ref{iso-contour-event}. 
This is in stark contrast to the case of hierarchy, where we saw that having data from two experiments with widely
different baselines is a necessity because the favorable and unfavorable combinations turned out to be
the same for $\nu$ and $\anu$.

Before discussing our results, we briefly describe the numerical procedure adopted.
We calculate $\dchsq$ using the default definition in GLoBES which is Poissonian.
We minimize this $\dchsq$ to compute the octant resolution capability. 
For a given octant-hierarchy combination and a true value of $\dcp$, 
we compute the events spectra and label it {\it data}.
Then we compute the theoretical events spectra where the octant is chosen to be the wrong one 
and the neutrino parameters are randomly chosen within
their allowed $3\sigma$ ranges.
We then calculate the $\dchsq$ between the data and the theoretical spectra.
This calculation uses the $\dchsq$ defined in GLoBES, which is valid
for a Poissonian distribution. We add a $\dchsq$ coming from gaussian priors on $|\Delta m^2_{\textrm{eff}}|$ and on
$\sin^22\ty$. The systematic uncertainties are included using the method of pulls.
If the $\dchsq_{\textrm{min}}\geq4$, 
then we can say that the wrong octant is ruled out at $2\sigma$ for the given combination and the given true value of
$\dcp$. This calculation is repeated for all true values of $\dcp$ and for all combinations. If the
$\dchsq_{\textrm{min}}\geq4$ in each case then the wrong octant can be ruled out independently of true
$\dcp$, hierarchy and octant.

\section{Results}
\label{results}

In this section, through various plots, we show the sensitivity of T2K and \nova
to the octant of $\theta_{23}$.

\subsection{Allowed regions in test $\sin^2\tz$ - true $\dcp$ plane}
\label{allowedtheta23plots}

\begin{figure}[H]

        \begin{tabular}{lr}
                \hspace*{-0.95in} \includegraphics[width=0.8\textwidth]
                {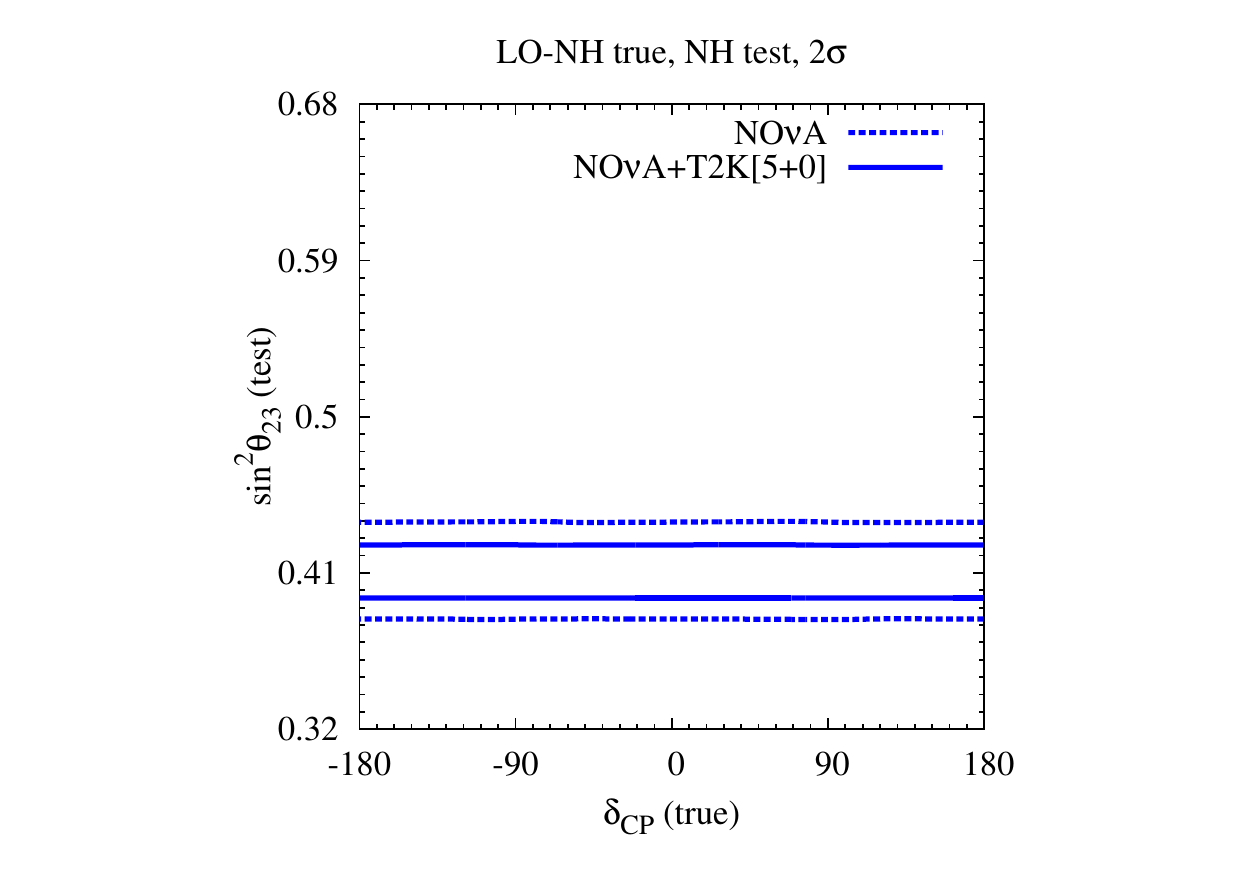}
                & 
                \hspace*{-2.0in} \includegraphics[width=0.8\textwidth]
                {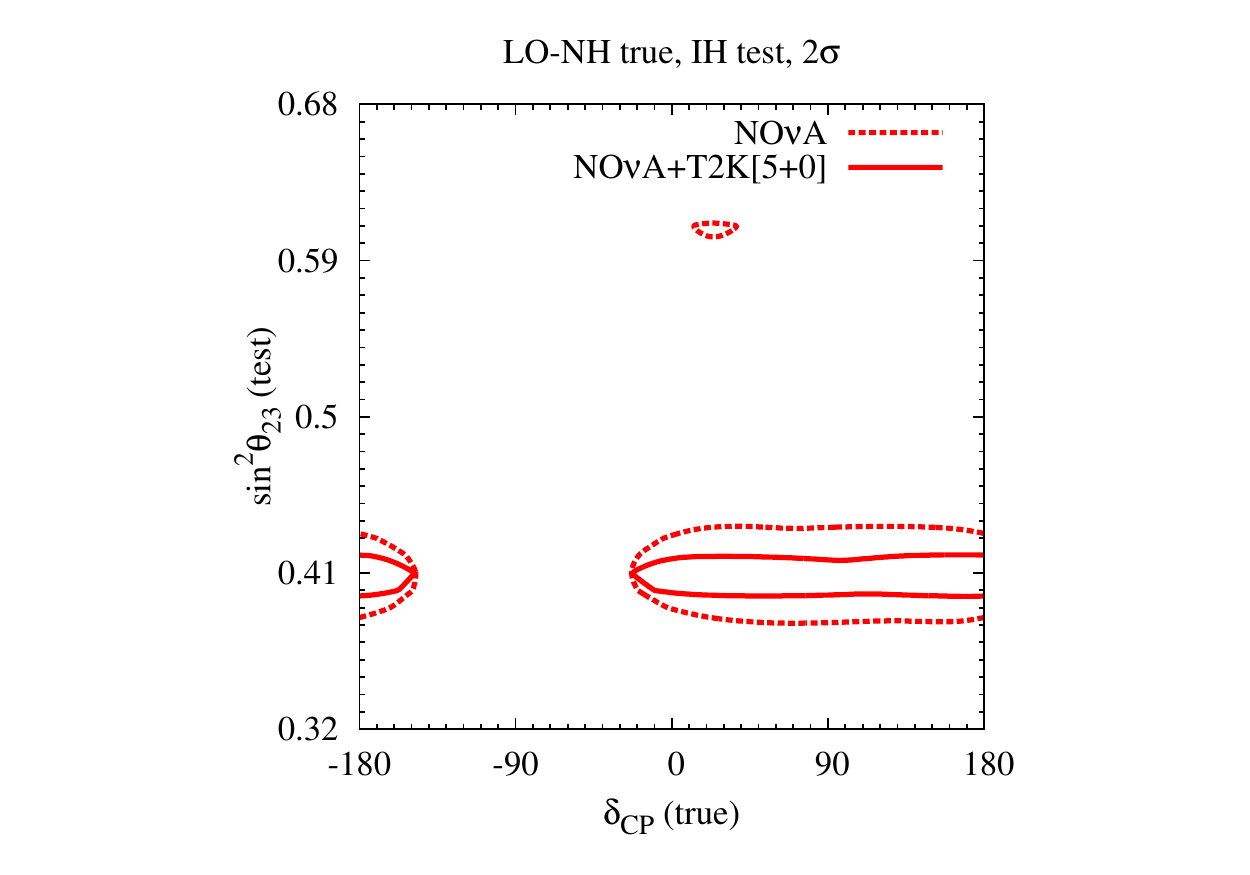}
        \end{tabular}
        
\caption{\label{allowedLONH-50}\footnotesize Allowed values of test $\sin^2\tz$ at $2\sigma$ (1 d.o.f.) C.L. 
as a function of true $\dcp$. LO-NH is assumed to be the true octant-hierarchy
combination. The left (right) panel corresponds to NH (IH) being the test hierarchy.}

\end{figure}

\begin{figure}[H]

        \begin{tabular}{lr}
                \hspace*{-0.95in} \includegraphics[width=0.8\textwidth]
                {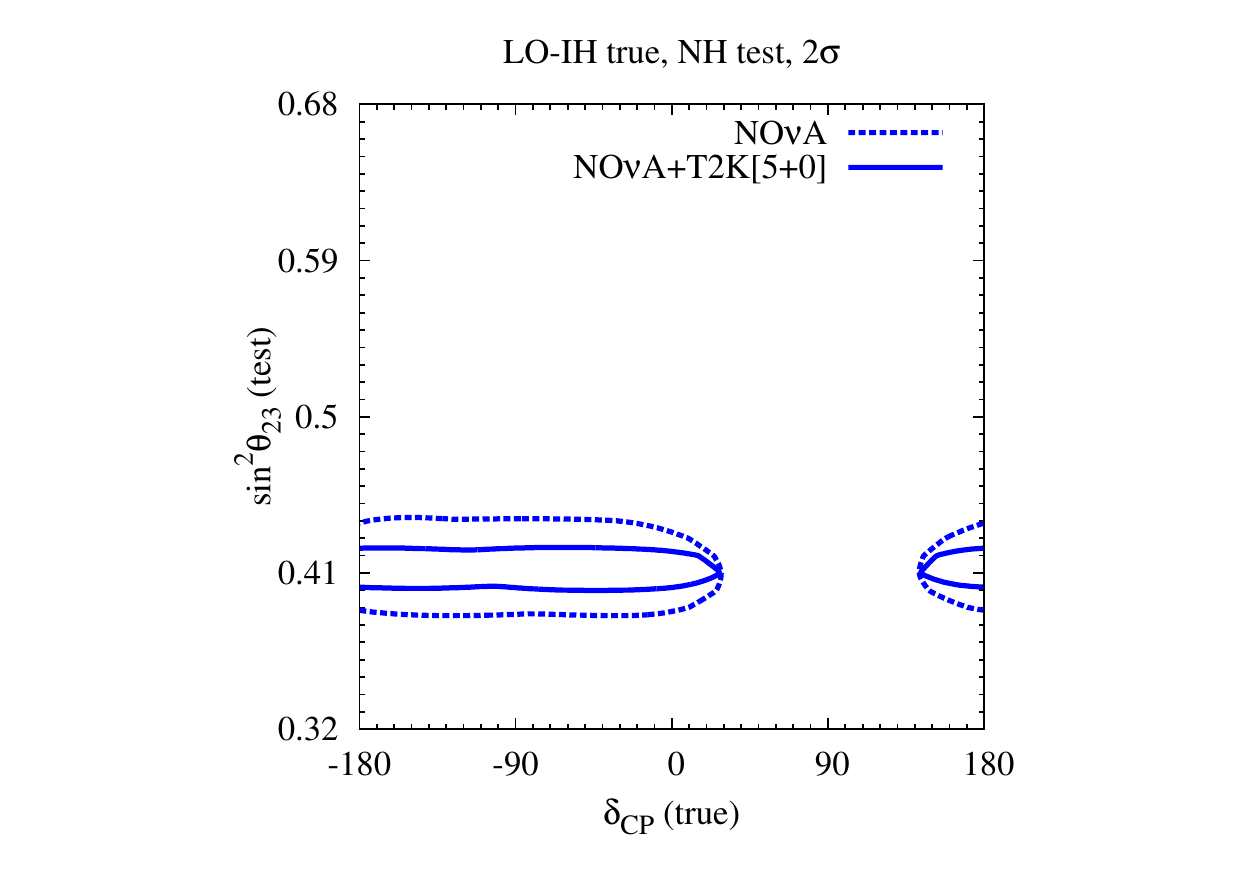}
                & 
                \hspace*{-2.0in} \includegraphics[width=0.8\textwidth]
                {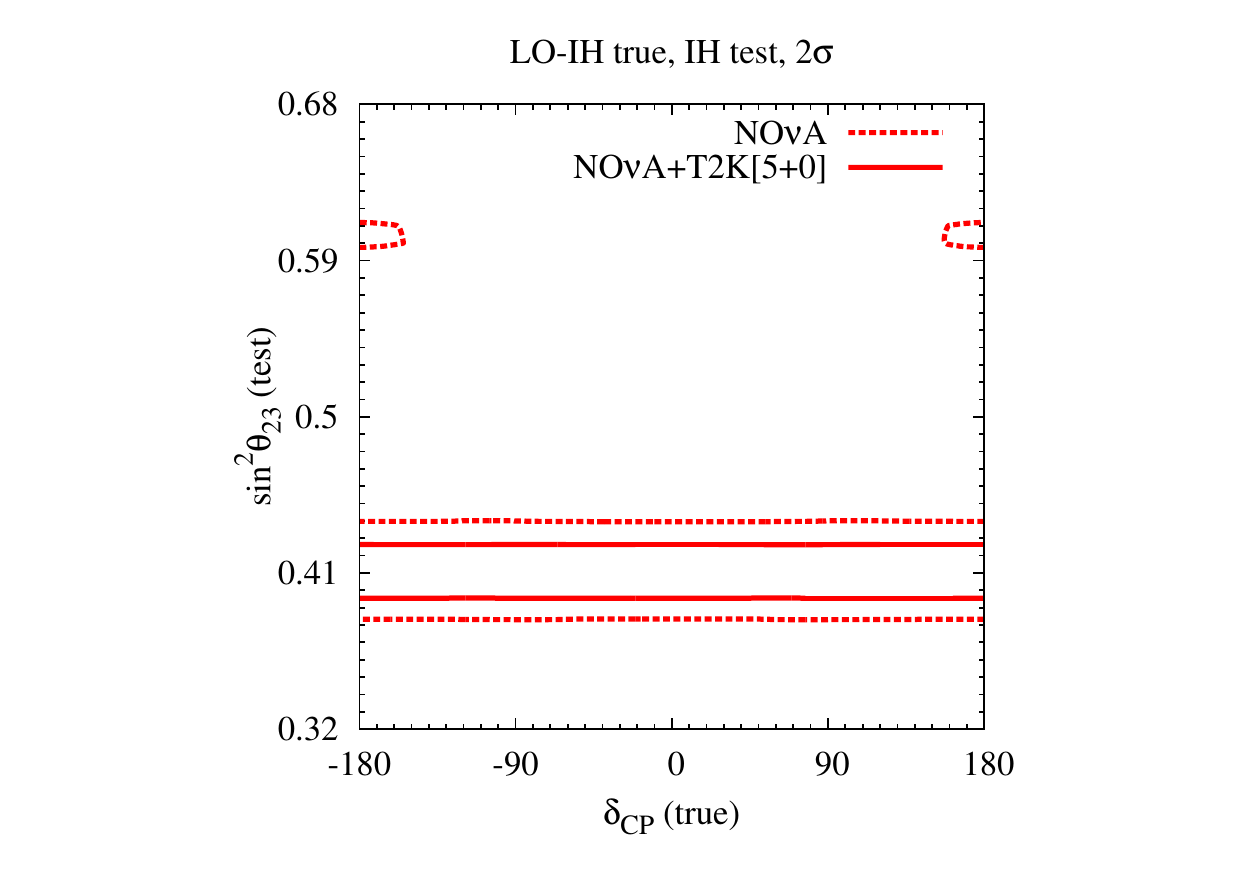}
        \end{tabular}
        
\caption{\label{allowedLOIH-50}\footnotesize Allowed values of test $\sin^2\tz$ at $2\sigma$ (1 d.o.f.) C.L. 
as a function of true $\dcp$. LO-IH is assumed to be the true octant-hierarchy
combination. The left (right) panel corresponds to NH (IH) being the test hierarchy.}

\end{figure}

\begin{figure}[H]

        \begin{tabular}{lr}
                \hspace*{-0.95in} \includegraphics[width=0.8\textwidth]
                {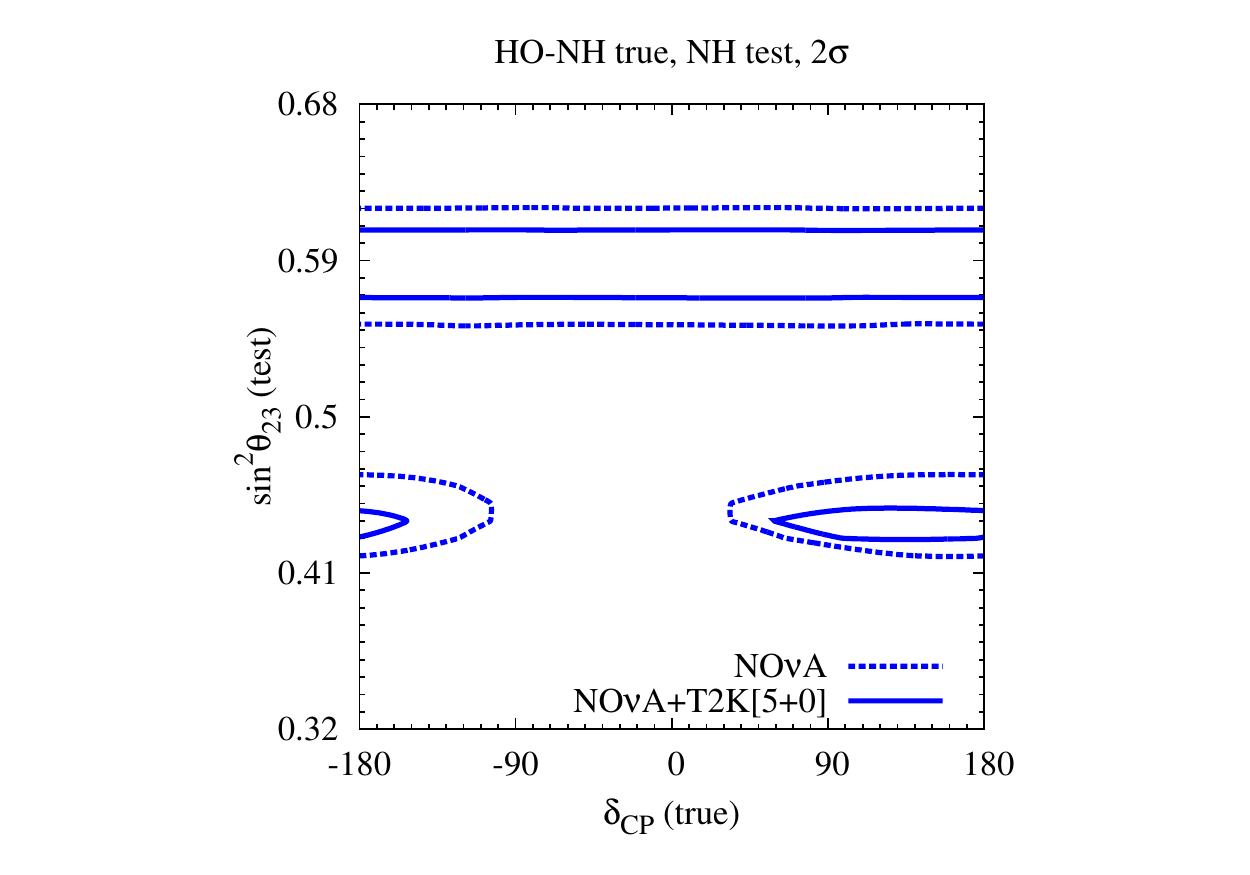}
                & 
                \hspace*{-2.0in} \includegraphics[width=0.8\textwidth]
                {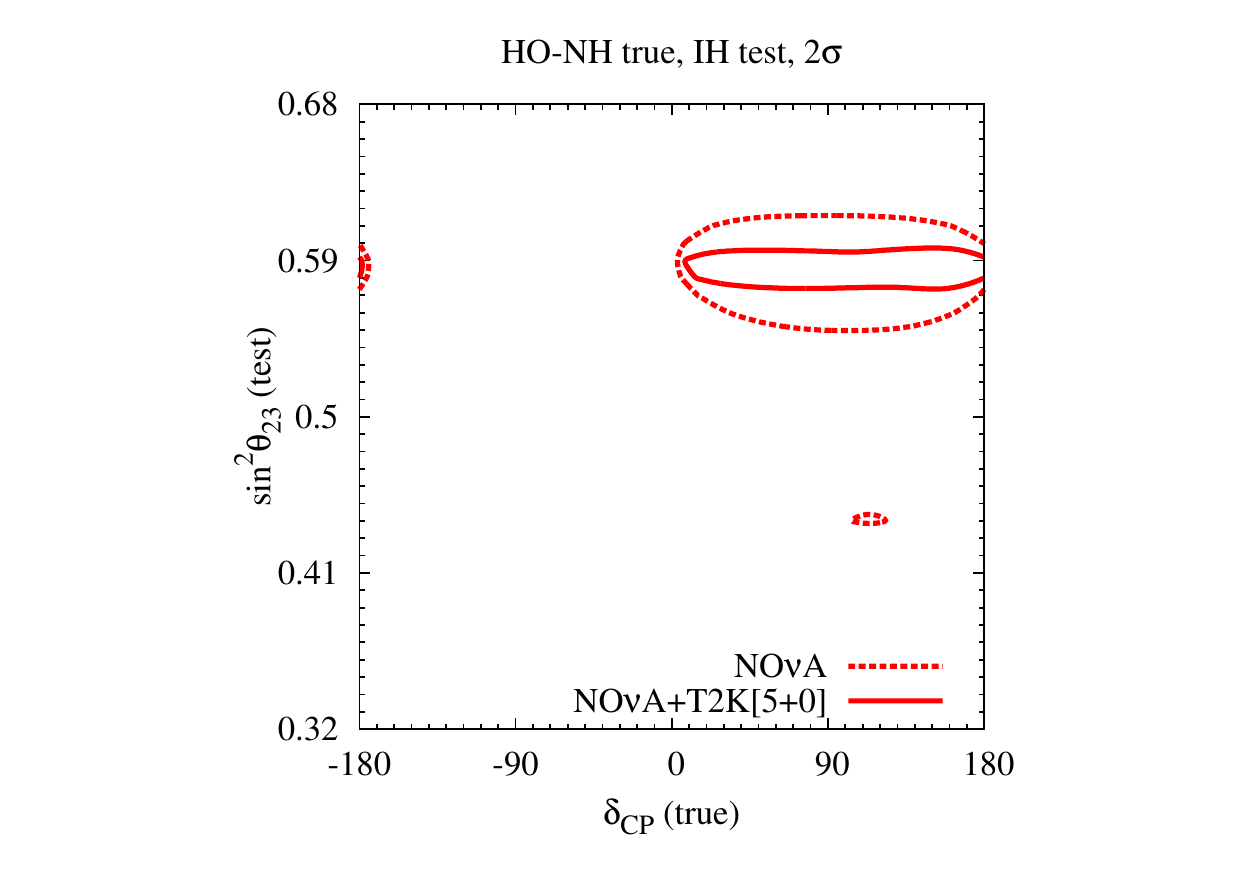}
        \end{tabular}
        
\caption{\label{allowedHONH-50}\footnotesize Allowed values of test $\sin^2\tz$ at $2\sigma$ (1 d.o.f. ) C.L. 
as a function of true $\dcp$. HO-NH is assumed to be the true octant-hierarchy
combination. The left (right) panel corresponds to NH (IH) being the test hierarchy.}

\end{figure}

\begin{figure}[H]

        \begin{tabular}{lr}
                \hspace*{-0.95in} \includegraphics[width=0.8\textwidth]
                {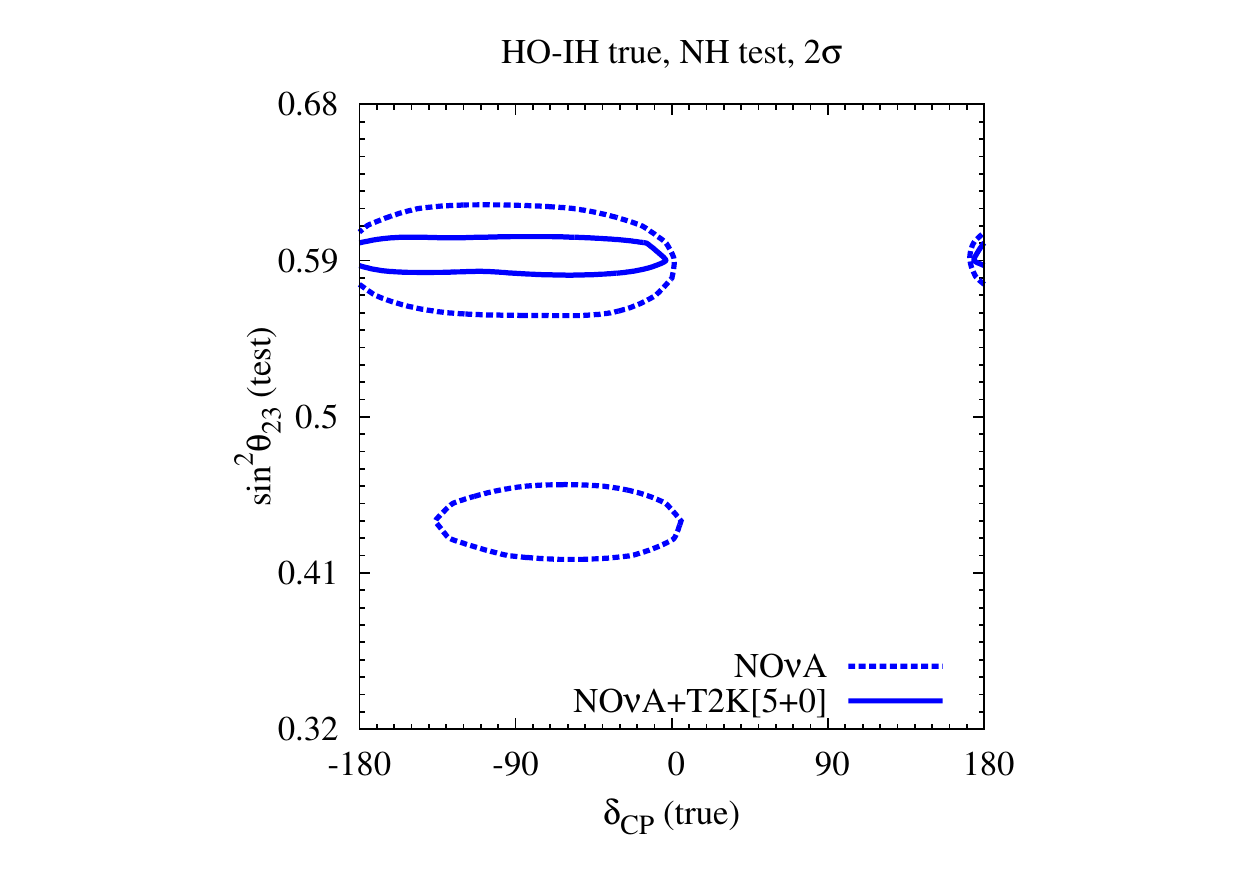}
                & 
                \hspace*{-2.0in} \includegraphics[width=0.8\textwidth]
                {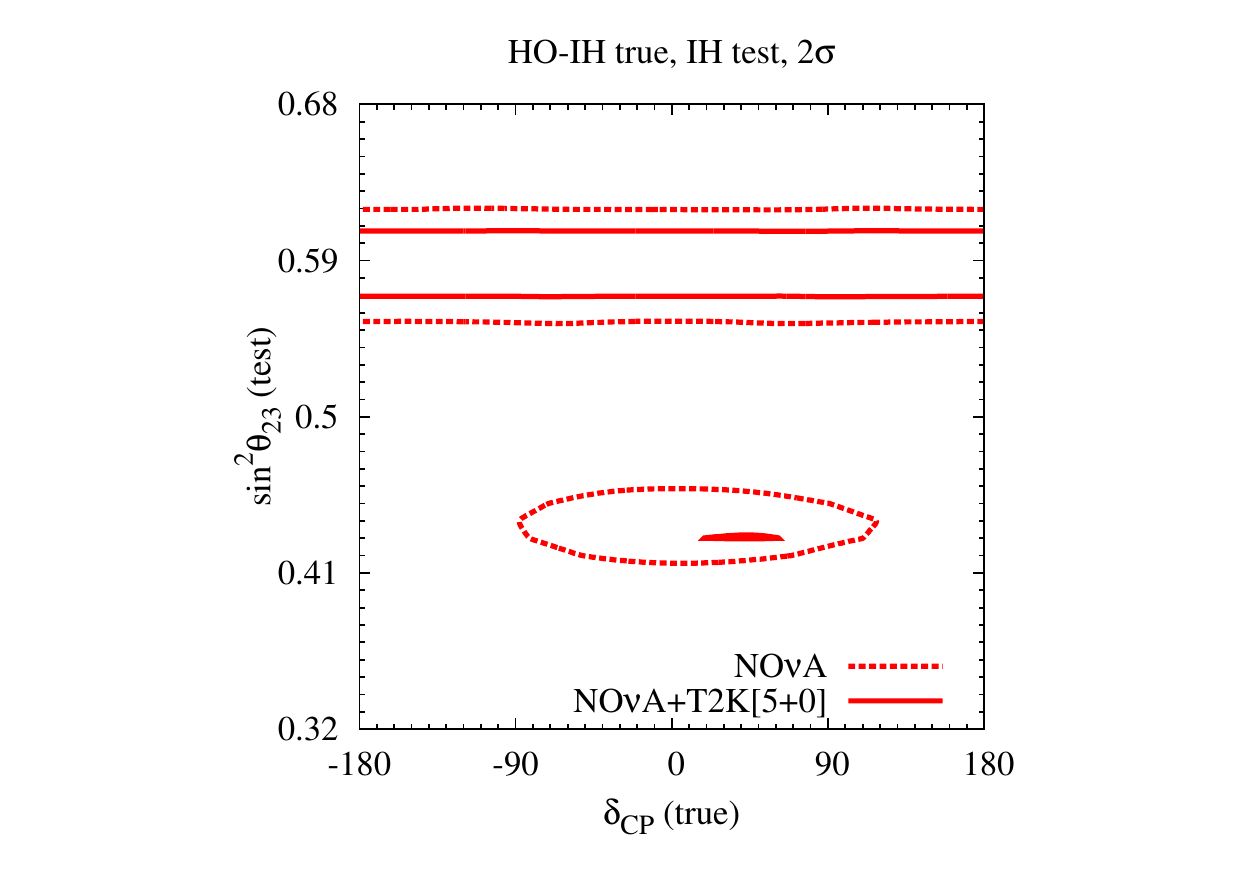}
        \end{tabular}
        
\caption{\label{allowedHOIH-50}\footnotesize Allowed values of test $\sin^2\tz$ at $2\sigma$ (1 d.o.f. ) C.L. 
as a function of true $\dcp$. HO-IH is assumed to be the true octant-hierarchy
combination. The left (right) panel corresponds to NH (IH) being the test hierarchy.}

\end{figure}

In figures \ref{allowedLONH-50}-\ref{allowedHOIH-50}, we have plotted the values of
test $\sin^2\tz$ allowed by T2K and \nova data as a function of true $\dcp$
for each of the four combinations of octant and hierarchy. In our calculations, we do not
assume a prior knowledge of hierarchy and hence consider both the possibilities for test hierarchy. 
Octant can be determined only if the wrong octant values are ruled out for both possibilities.

In this analysis, we have included the data from both the disappearance channel $\textrm{P}_{\mu \mu}$
and the appearance channel $\textrm{P}_{\mu e}$.
We varied $\sin^2\tz$ within its $3\sigma$
range: [0.34, 0.67] allowed by the current global fits.
Here, we are constraining $\sin^2\tz$ only and are marginalizing over $\dcp$
and hence take the $2\sigma$ limit relevant for 1 d.o.f. The contours in these figures 
are defined by $\Delta \chi^2 \leq 4$. 
{\it The disappearance channel gives a precise measurement of $\sin^22\theta_{\textrm{eff}}$ which
leads to two narrow, $\dcp$-independent, allowed $\sin^2\tz$ bands, one in each octant. 
The appearance channel, because of its large octant sensitivity, 
discriminates against the wrong octant band}. If the statistics are large enough, the 
wrong octant can be ruled out.

\begin{figure}[H]

        \begin{tabular}{lr}
                \hspace*{-0.95in} \includegraphics[width=0.8\textwidth]
                {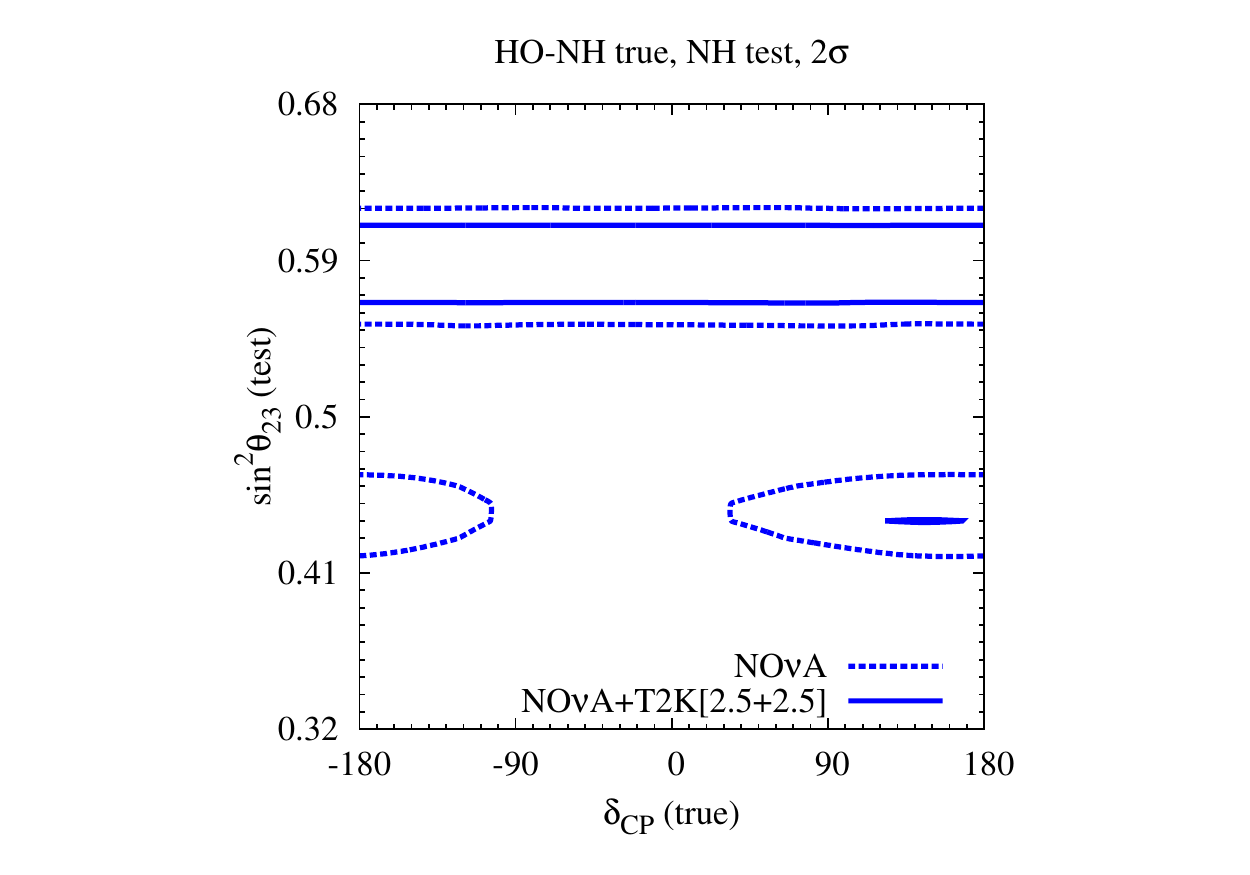}
                & 
                \hspace*{-2.0in} \includegraphics[width=0.8\textwidth]
                {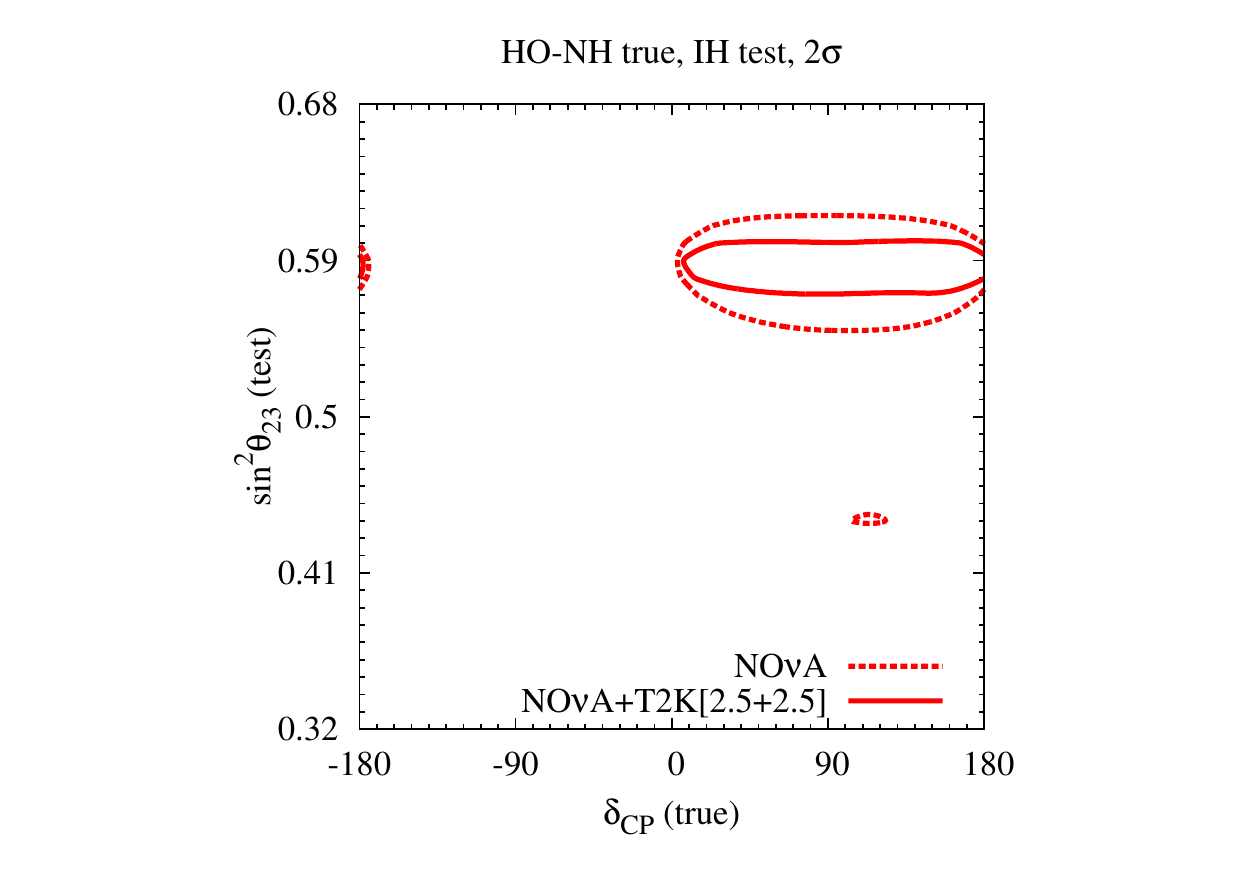}
        \end{tabular}
        
\caption{\label{allowedHONH-2525}\footnotesize Allowed values of test $\sin^2\tz$ at $2\sigma$ (1 d.o.f. ) C.L. 
as a function of true $\dcp$. HO-NH is assumed to be the true octant-hierarchy
combination. The left (right) panel corresponds to NH (IH) being the test hierarchy.
Note that for T2K, equal $\nu$ and $\anu$ runs of 2.5 years each has been assumed.}

\end{figure}

From figures \ref{allowedLONH-50} and \ref{allowedLOIH-50}, we note that the combined data of
T2K and \nova can rule out HO if LO is the true octant. This holds for both NH and IH.
However, if HO is the true octant, then LO can not be ruled 
out for a reasonably large fraction of true $\dcp$.

\begin{figure}[H]

        \begin{tabular}{lr}
                \hspace*{-0.95in} \includegraphics[width=0.8\textwidth]
                {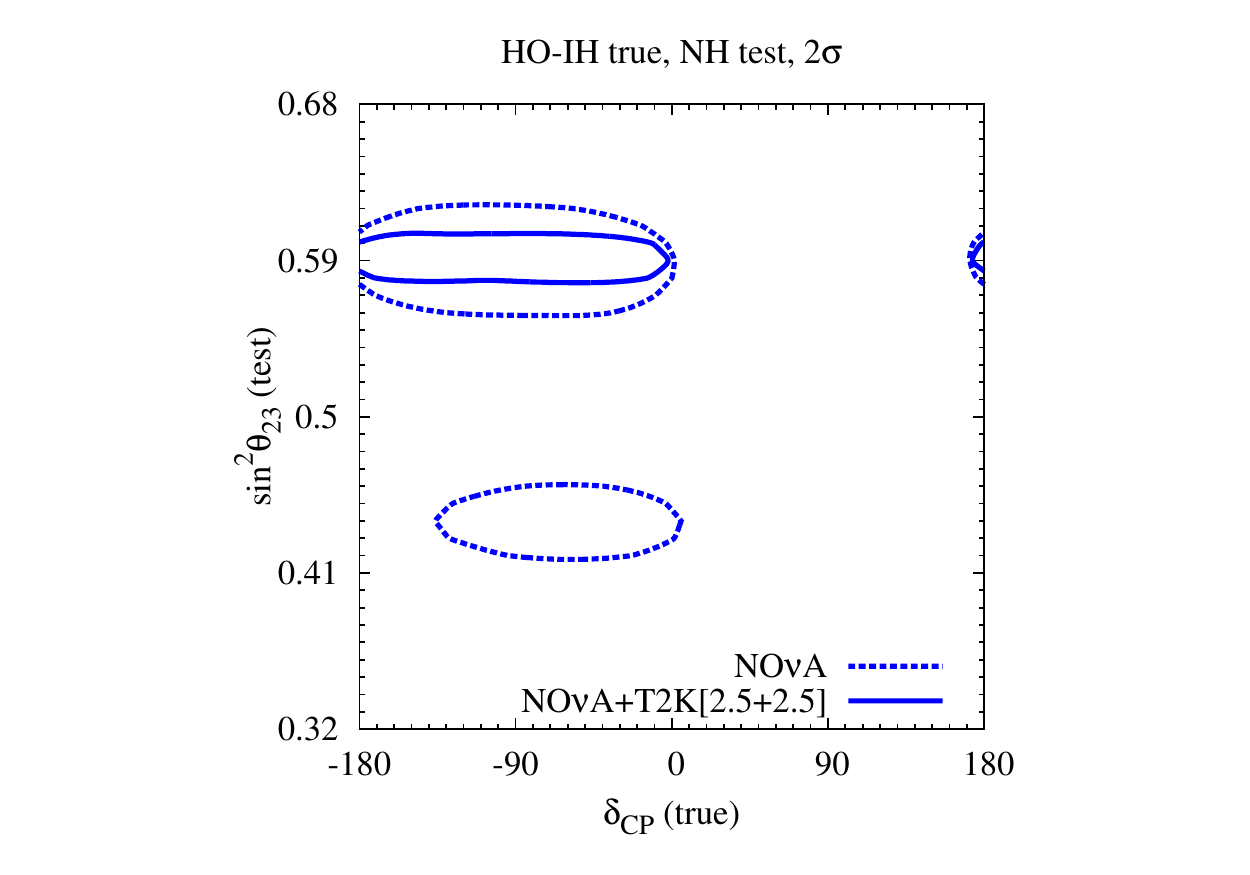}
                & 
                \hspace*{-2.0in} \includegraphics[width=0.8\textwidth]
                {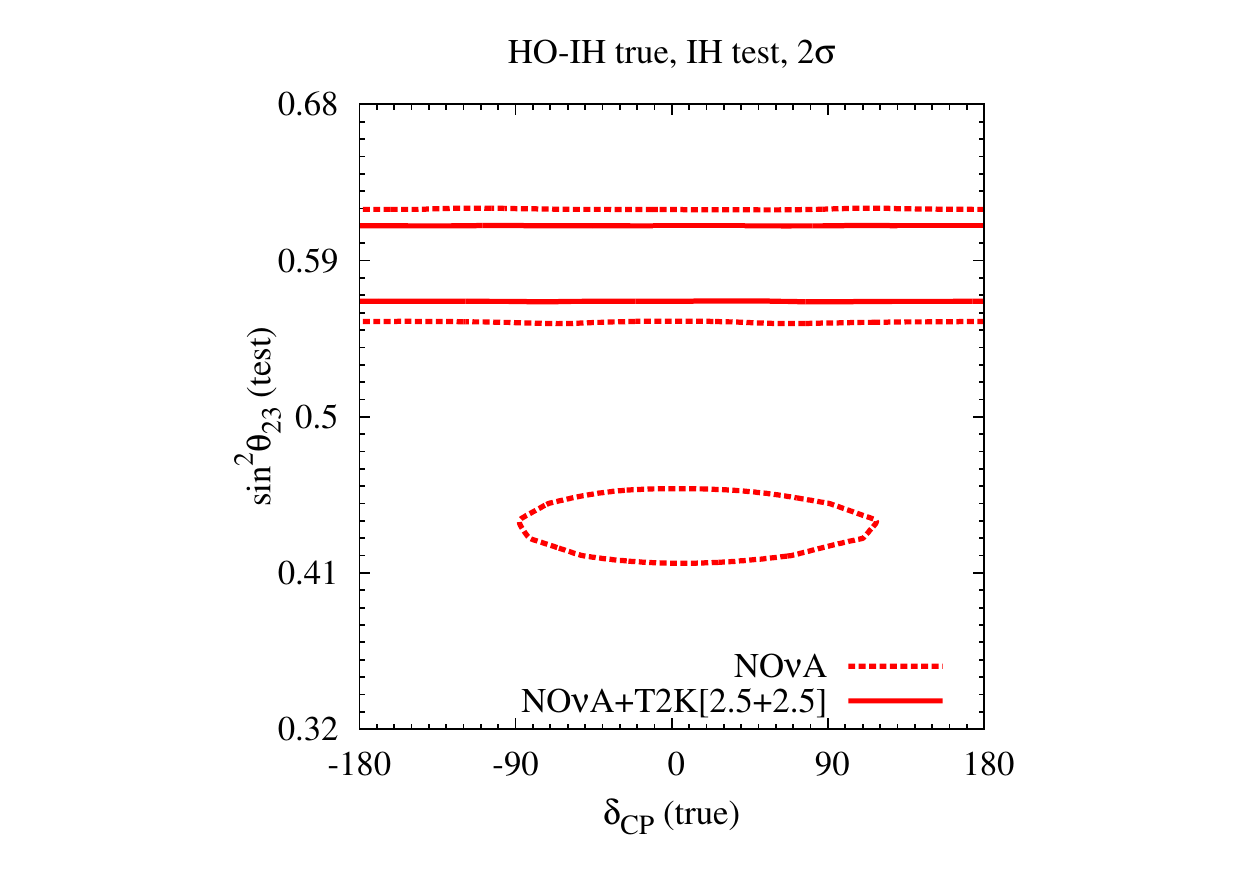}
        \end{tabular}
        
\caption{\label{allowedHOIH-2525}\footnotesize Allowed values of test $\sin^2\tz$ at $2\sigma$ (1 d.o.f. ) C.L. 
as a function of true $\dcp$. HO-IH is assumed to be the true octant-hierarchy
combination. The left (right) panel corresponds to NH (IH) being the test hierarchy.
Note that for T2K, equal $\nu$ and $\anu$ runs of 2.5 years each has been assumed.}

\end{figure}

As argued in section \ref{physics}, the favorable and unfavorable regions are very different
for neutrinos and anti-neutrinos. Hence, we explored if an improvement in the octant determination
can be achieved if T2K has equal neutrino and anti-neutrino runs of 2.5 years 
each\footnote{In GLoBES, the background events for T2K are given for a 5 year $\nu$ run
taken from \cite{fechnerthesis}. We have taken care to do appropriate scaling of these events
in computing for ($2.5\nu+2.5\anu$) runs.}.
As expected, in the two cases of LO-NH and LO-IH, a ($2.5\nu+2.5\anu$) 
runs in T2K is also effective in ruling out the wrong octant at $2\sigma$. Hence, we have not displayed the
corresponding figures. But, for the two cases HO-NH and HO-IH, the balanced $\nu$ and $\anu$ runs are more effective
than the ($5\nu$) run. This is shown in figures \ref{allowedHONH-2525} and \ref{allowedHOIH-2525}.
In the left panel of figure \ref{allowedHONH-2525}, there is a very small sliver of allowed 
$\sin^2\tz$ in the wrong octant for true $\dcp$ in the range $[130^\circ,160^\circ]$. But, 
the $\Delta\chi^2$ for this sliver is very close to 4. Hence, this wrong octant region can
be effectively discriminated against, leading to a determination of the octant of $\tz$
for any true value of $\dcp$.

The above set of figures also give the precision on $\sin^2\tz$ that one can obtain. 
This precision is the result of the precise measurement of $\sin^22\tz$ from the disappearance channel.
From figures \ref{allowedLONH-50} and \ref{allowedLOIH-50}, we get $\delta(\sin^2\tz)=0.015$ if LO is
the true octant. From figures \ref{allowedHONH-50} and \ref{allowedHOIH-50}, 
we see that a 5 year $\nu$ run of T2K combined with \nova gives $\delta(\sin^2\tz)=0.02$ if HO is
the true octant. When the T2K run is changed to ($2.5\nu+2.5\anu$), the precision in $\sin^2\tz$ becomes
slightly worse because of the loss of statistics caused by $\anu$ run as seen in figures 
\ref{allowedHONH-2525} and \ref{allowedHOIH-2525}.

\subsection{$\Delta\chi^2$ vs. true $\dcp$ plots}
\label{exclusionplots}

\begin{figure}[H]

        \begin{tabular}{lr}
                \hspace*{-0.95in} \includegraphics[width=0.8\textwidth]
                {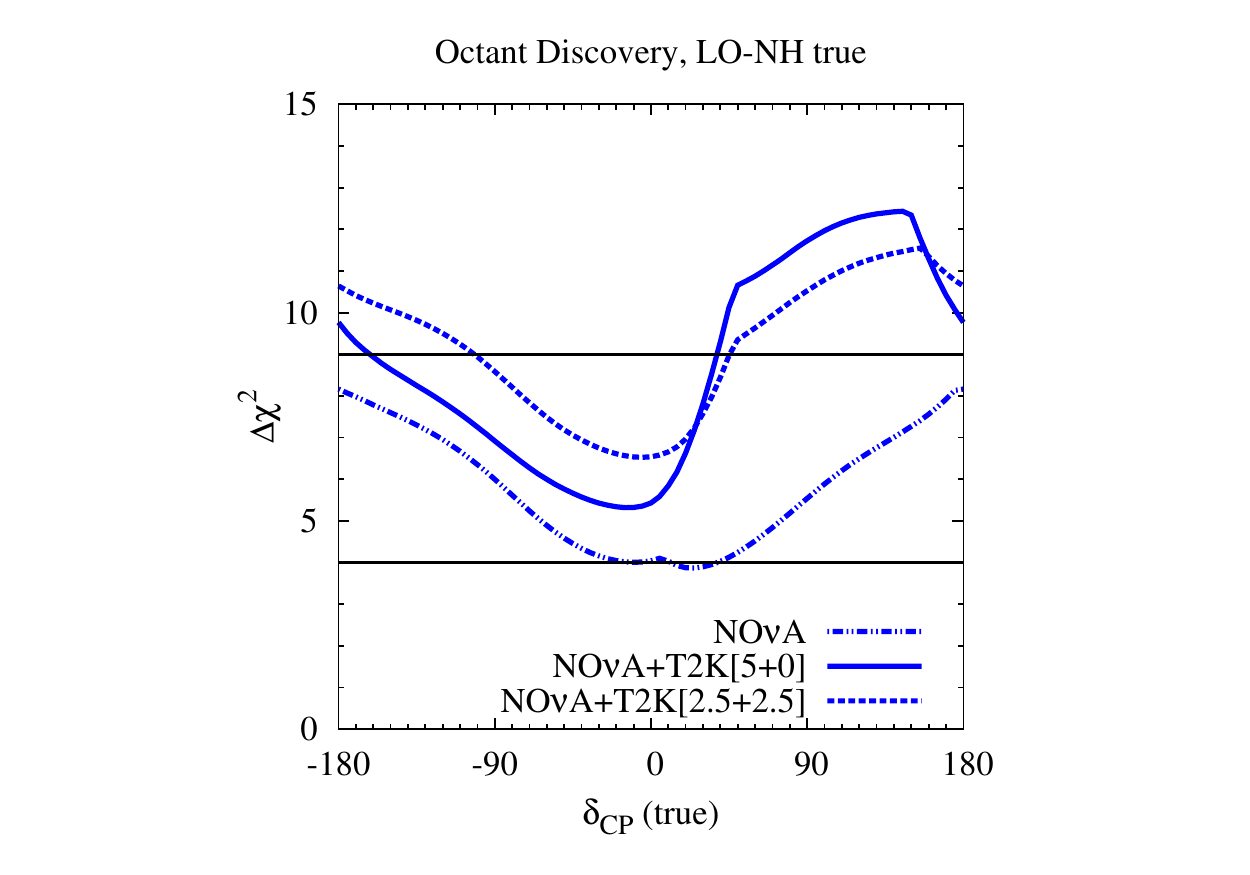}
                & 
                \hspace*{-2.1in} \includegraphics[width=0.8\textwidth]
                {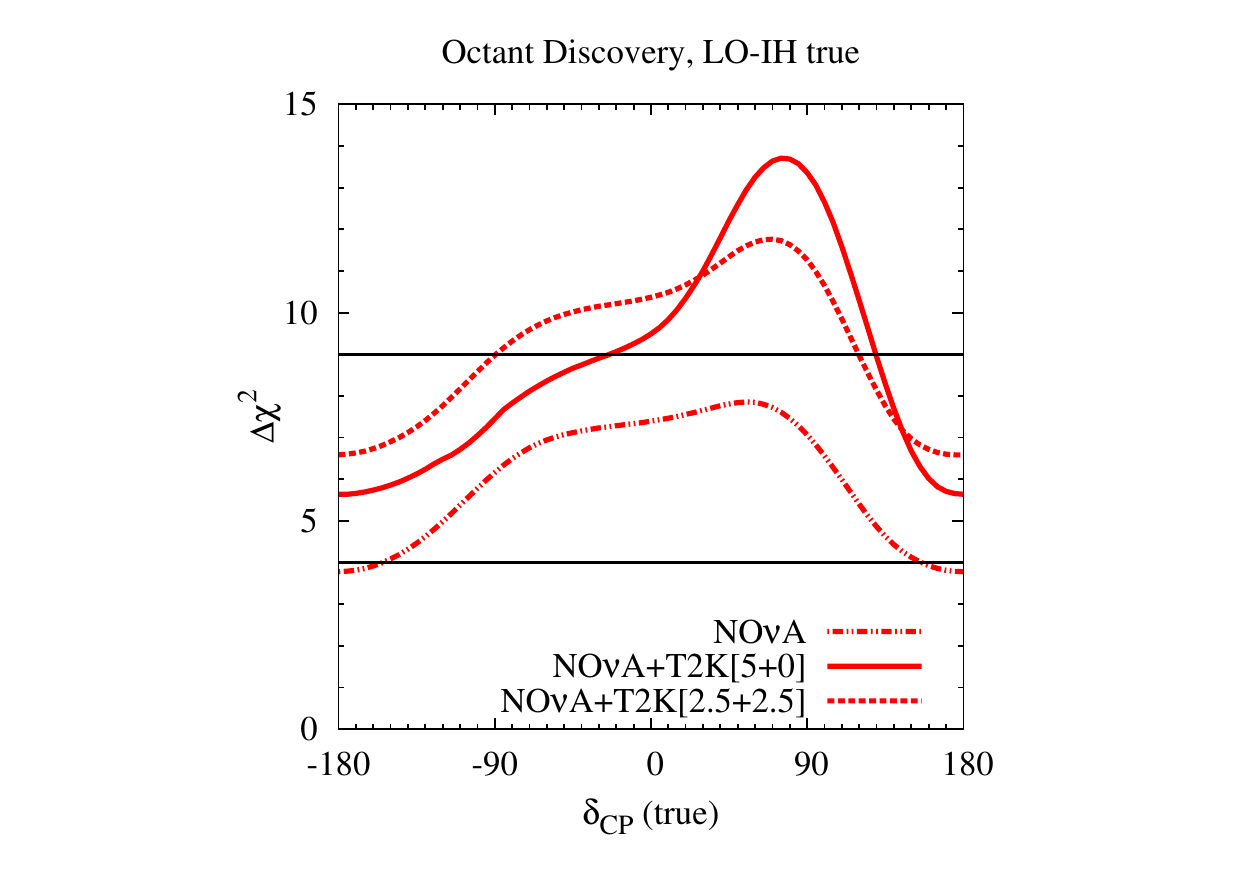}
        \end{tabular}
        
\caption{\label{dchsqLO}\footnotesize Octant resolving capability as a function 
of true $\dcp$ for various set-ups . In these plots, LO is assumed to be
the true octant. The left (right) panel corresponds to NH (IH) being the true hierarchy.}

\end{figure}

\begin{figure}[H]

        \begin{tabular}{lr}
                \hspace*{-0.95in} \includegraphics[width=0.8\textwidth]
                {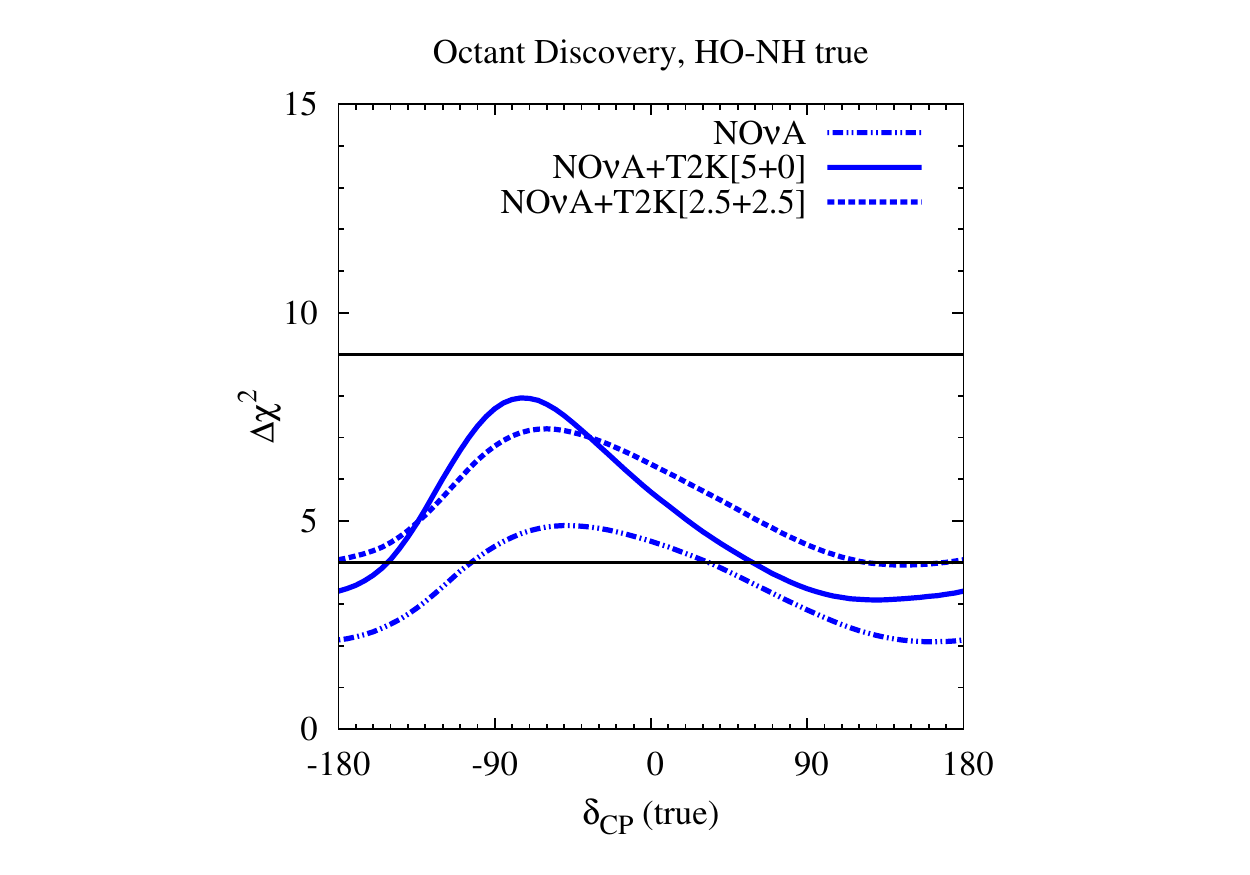}
                & 
                \hspace*{-2.1in} \includegraphics[width=0.8\textwidth]
                {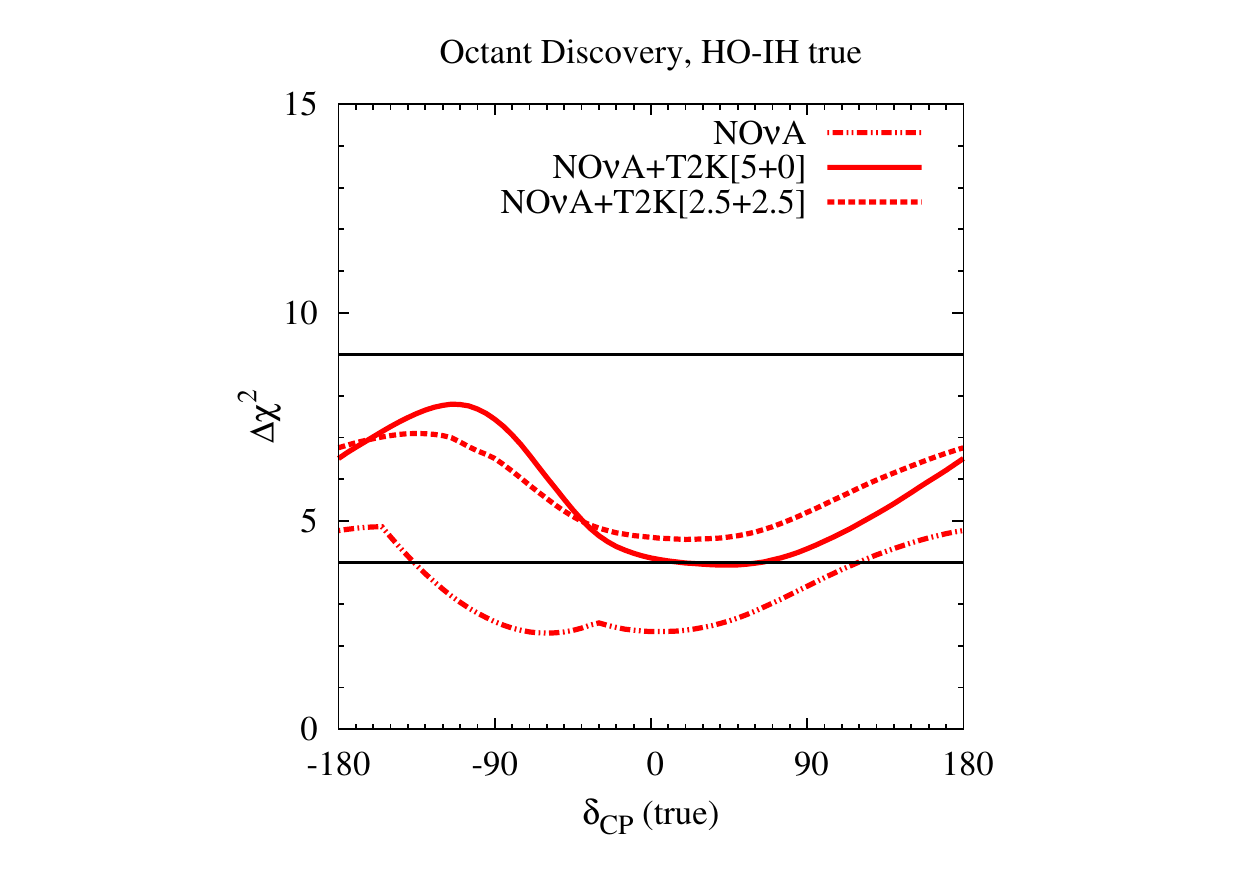}
        \end{tabular}
        
\caption{\label{dchsqHO}\footnotesize Octant resolving capability as a function 
of true $\dcp$ for various set-ups . In these plots, HO is assumed to be
the true octant. The left (right) panel corresponds to NH (IH) being the true hierarchy.}

\end{figure}

In this section, we study the behavior of $\Delta\chi^2$ between 
the true and the wrong octants as a function of true $\dcp$. 
Here, the $\dchsq$ is computed 
in the following way. First, we fix the true value of $\dcp$.
We take $\sin^2\tz$ to be its best-fit value in the true octant: 0.41 for LO and
0.59 for HO. If the LO (HO) is the true octant, the  
test values of $\sin^2\tz$ in the HO (LO) are varied within 
the range $[0.5,0.63]$ ($[0.36,0.5]$),
where 0.63 (0.36) is the $2\sigma$ upper (lower) limit of the allowed range of $\sin^2\tz$.
The $\dchsq$ is computed between the spectra with the best-fit $\sin^2\tz$ of the
true octant and that with various test values in the wrong octant and is marginalized over other neutrino
parameters, especially the hierarchy, $\sin^22\ty$ and $\dcp$. 
Figures \ref{dchsqLO} and \ref{dchsqHO} show the minimum of this $\dchsq$ vs. the true value of $\dcp$.

From figure \ref{dchsqLO}, we see that the \nova data by itself can almost rule out the wrong
octant at $2\sigma$, if LO is the true octant. We see that the $\dchsq$ dips just below 4 for  
true $\dcp \sim 0(180^\circ)$ if the true hierarchy is NH (IH). But, as argued earlier, this small allowed
region can be effectively discriminated because of the relatively large $\dchsq$. If HO is the
true octant, then \nova data is not sufficient to rule out the wrong octant as seen in figure \ref{dchsqHO}.
In fact, the wrong octant can be ruled out only for about half of the true $\dcp$ values. 
As illustrated in figures \ref{dchsqLO} and \ref{dchsqHO}, addition 
of T2K data improves the octant determination ability significantly. 

From figure \ref{dchsqLO}, we see that the combined data from \nova and T2K (5$\nu$) give a 2$\sigma$
octant resolution for all values of true $\dcp$ if LO is the true octant. From figure \ref{dchsqHO},
we see that this combined data can rule out the wrong octant at $2\sigma$ for HO-IH, but not for HO-NH.
The problem of HO-NH can be solved if the T2K has equal $\nu$ and $\anu$ runs of 2.5 years each.
This change improves the octant determination for the unfavorable values of true $\dcp$ (where
$\dchsq$ is minimum) for all four combinations of hierarchy and octant. In particular, for the case of HO-NH,
it leads to a complete ruling out of the wrong octant at $2\sigma$ for all values of true $\dcp$. Thus,
balanced runs of T2K in $\nu-\anu$ mode is preferred over a pure $\nu$ run because of better octant determination
capability.

Figures \ref{dchsqLO} and \ref{dchsqHO} show that the combined data from \nova and T2K has a 
better overall octant resolving capability if LO is the true octant. We found out that this 
feature of LO being more favorable compared to HO is a consequence of marginalization over 
the oscillation parameters (mainly $\dcp$) and the systematic uncertainties. We checked that in the absence 
of any kind of marginalization $\Delta\chi^2_{\textrm{HO}}$ is consistently larger than
$\Delta\chi^2_{\textrm{LO}}$.


\subsection{Octant resolution as function of true $\tz$}

In the discussion so far, we assumed the true values of $\sin^2\tz$ to be
0.41 for LO and 0.59 for HO. These are, of course, the best-fit points from the 
global analyses. But, we must consider the octant resolution capability
for values of true $\sin^2\tz$ in the full allowed range (0.34, 0.67). In this
subsection, we calculate the octant resolution capability of \nova and T2K
as a function of both true $\sin^2\tz$ and true $\dcp$. It was shown earlier that
equal $\nu$-$\anu$ runs are superior to pure $\nu$ run for octant resolution. 
Hence, in this subsection, the calculations are done only for the following
run combination: ($3\nu+3\anu$) for \nova and ($2.5\nu+2.5\anu$) for T2K.

\begin{figure}[H]

        \begin{tabular}{lr}
                \hspace*{-0.95in} \includegraphics[width=0.8\textwidth]
                {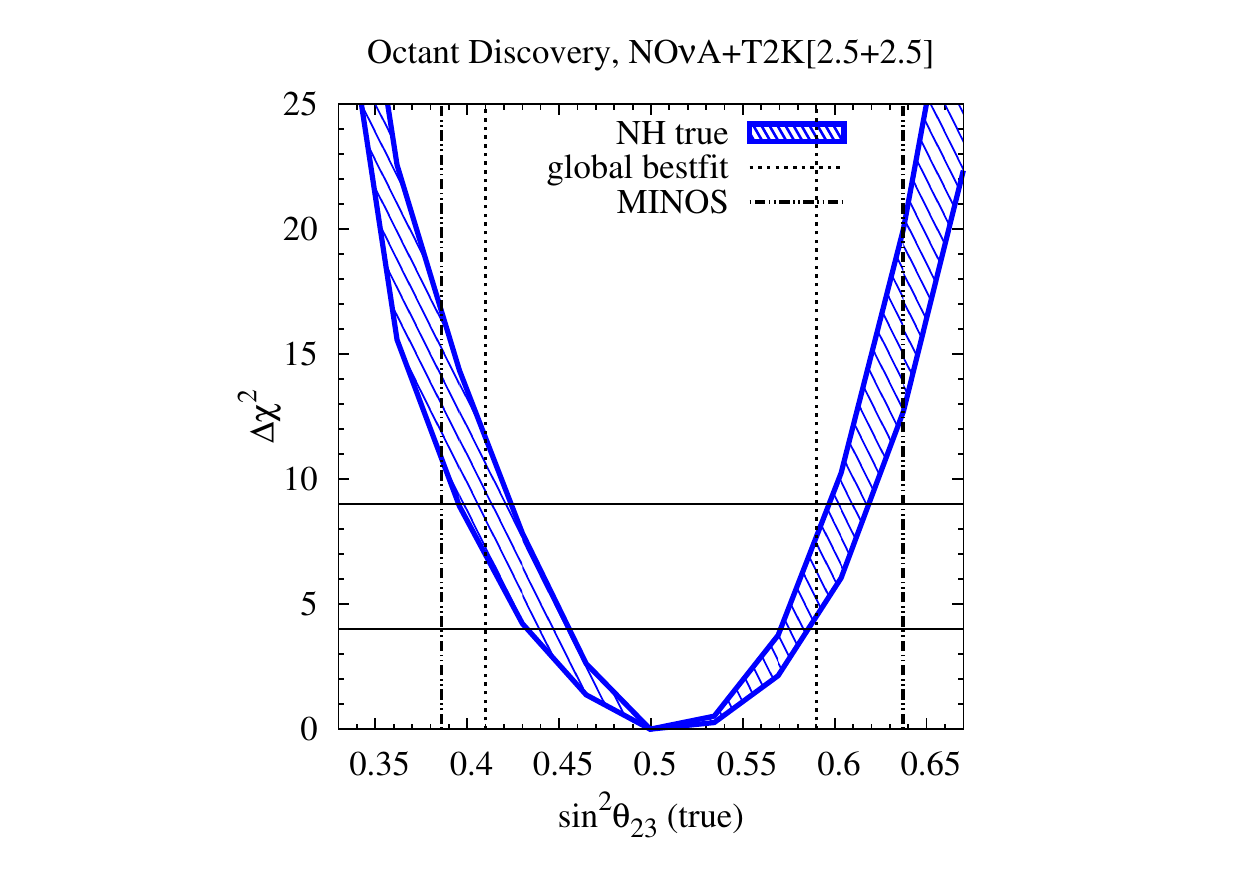}
                & 
                \hspace*{-2.1in} \includegraphics[width=0.8\textwidth]
                {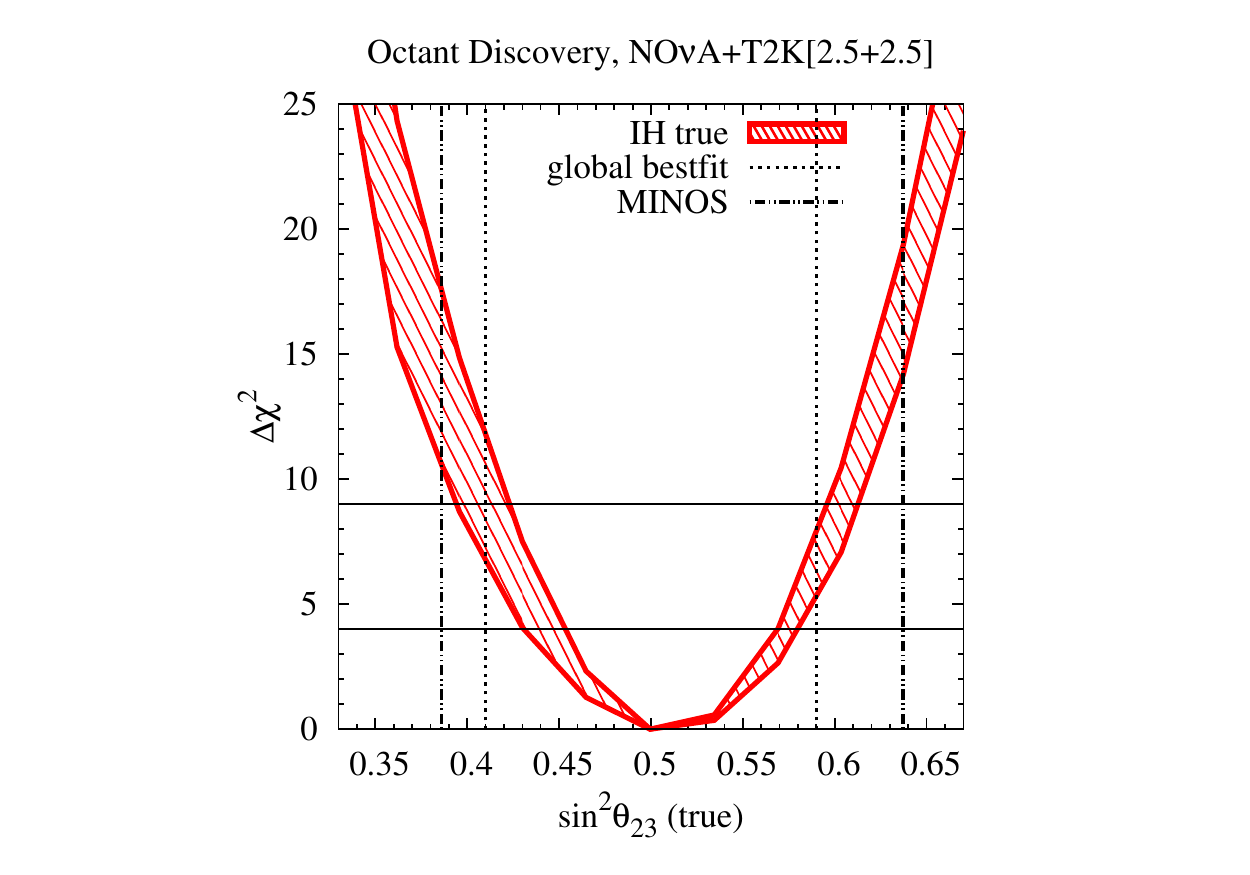}
        \end{tabular}
        
\caption{\label{dchsq_true_tz}\footnotesize Octant resolving capability as a function 
of true $\sin^2\tz$ for the combined $3\nu+3\anu$ runs of \nova and $2.5\nu+2.5\anu$ runs of T2K. 
The variation in values of $\Delta\chi^2$ for a given true $\sin^2\tz$ is due
to the variation in true $\dcp$. The vertical lines correspond to the
best-fit values of global data and those of MINOS accelerator data.
The left (right) panel corresponds to NH (IH) being the true hierarchy.}

\end{figure}

In figure \ref{dchsq_true_tz}, we plotted the $\Delta\chi^2$ vs. the true value of $\sin^2\tz$.
This $\Delta\chi^2$ is computed for the given true value of $\sin^2\tz$ with true $\dcp$ varying
between (-180$^\circ$,180$^\circ$). This variation in $\dcp$ leads to the band of values 
in $\Delta\chi^2$. As expected, the octant resolution is poor for true $\sin^2\tz$ close to 0.5.
A $2\sigma$ octant resolution is possible for $\sin^2\tz\leq0.43$ and for $\sin^2\tz\geq0.58$,
for all values of $\dcp$. This is consistent with our earlier claim that $2\sigma$ octant 
resolution is possible for the global best-fit values of $\sin^2\tz$.

As mentioned earlier, MINOS favors a non-maximal value of $\sin^22\theta_{\textrm{\textrm{eff}}}=0.94$. 
This gives two degenerate solutions, $\tz\approx38^\circ$ ($\sin^2\tz=0.386$) in LO and 
$\tz\approx53^\circ$ ($\sin^2\tz=0.637$) in HO. Note that these values are not complementary
because of the $\ty$-dependent correction described in equation \ref{tzeff}. For these values, a better than $3\sigma$
octant resolution is possible, which is true for both choices of true hierarchy.

\begin{figure}[H]

        \begin{tabular}{lr}
                \hspace*{-0.95in} \includegraphics[width=0.8\textwidth]
                {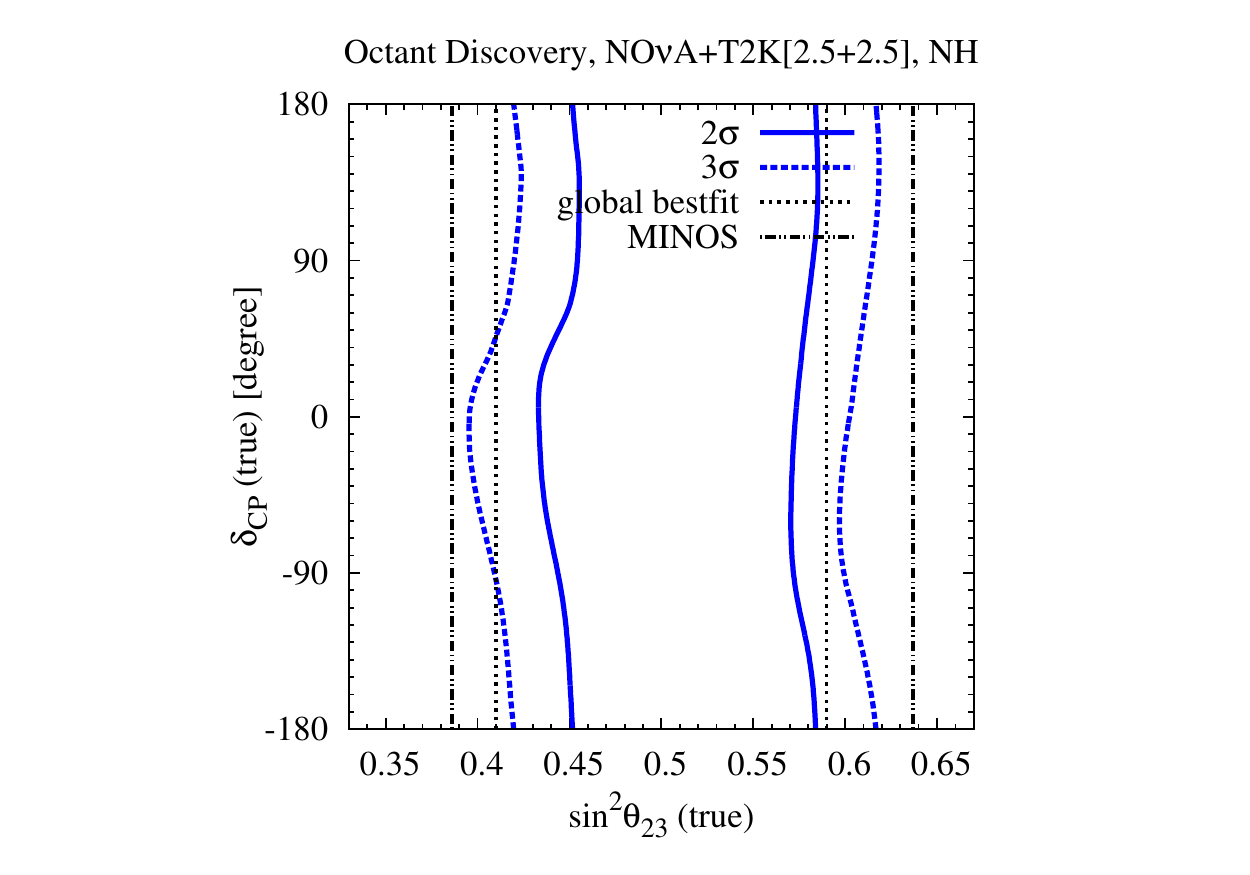}
                & 
                \hspace*{-2.1in} \includegraphics[width=0.8\textwidth]
                {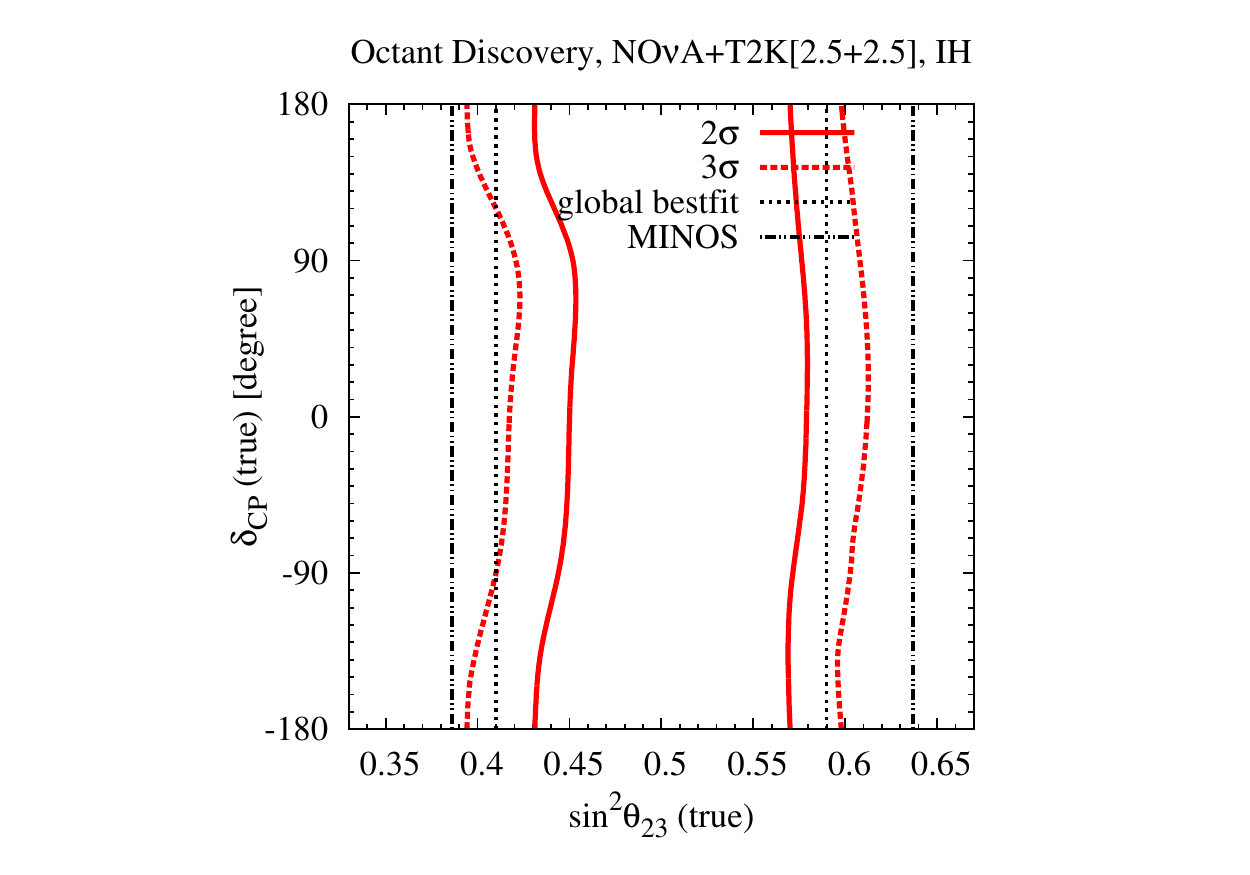}
        \end{tabular}
        
\caption{\label{true_tz_true_dcp}\footnotesize Octant resolving capability in the  
true $\sin^2\tz$ - true $\dcp$ plane for the combined $3\nu+3\anu$ runs of 
\nova and $2.5\nu+2.5\anu$ runs of T2K. Both $2\sigma$ and $3\sigma$
C.L. contours are plotted.
The vertical lines correspond to the
best-fit values of global data and those of MINOS accelerator data.
The left (right) panel corresponds to NH (IH) being the true hierarchy.}

\end{figure}

In figure \ref{true_tz_true_dcp}, we plotted the $2\sigma$ and $3\sigma$ octant
resolution contours in true $\sin^2\tz$ - true $\dcp$ plane. Octant resolution 
is possible only for points lying outside the contours. 
These figures again show that octant resolution
is possible at $2\sigma$ for global best-fit points and at $3\sigma$ for MINOS
best-fit points. The results for the two hierarchies are similar here also.

\section{Summary and Conclusions}
\label{summary}

Recently, the preliminary results from MINOS experiment have indicated that $\tz$ is not maximal. 
This raises the question of the true octant of $\tz$ \ie~whether $\tz < 45^\circ$ (LO)
or $\tz > 45^\circ$ (HO). $\nu_{e}$ appearance searches at the presently running T2K and upcoming \nova
experiments are sensitive to the octant, especially in light of moderately large $\ty$.
The main difficulty in octant resolution stems from octant-$\dcp$ degeneracy.
We explored this degeneracy in detail and found that equal neutrino and anti-neutrino
runs are mandatory to overcome this problem. 
This is because the octant-$\dcp$ combinations, which have degenerate
probabilities in the neutrino data are free of this degeneracy in the anti-neutrino data and vice-verse.
We observe that equal $\nu-\anu$ runs for T2K and \nova overcome the octant-$\dcp$
degeneracy for both NH and IH.
We also studied the prospects of T2K and \nova to improve the precision in $\tz$. 
The $\nu_{\mu}$ disappearance
data leads to two degenerate solutions of $\sin^2\tz$, one in each octant.
For $\sin^2\tz=0.41~(0.59)$, the expected precision will be $\delta(\sin^2\tz)= 0.015~(0.02)$.
The $\nu_e$ appearance data can resolve the octant degeneracy for the best-fit values. \nova alone can rule 
out the wrong octant at $2\sigma$ if $\sin^2\tz=0.41$, 
independently of hierarchy and $\dcp$. 
Combined data from equal neutrino and anti-neutrino runs of T2K (2.5 years each) and 
\nova (3 years each) can establish the correct octant at $2\sigma$ C.L. for any combination of hierarchy and $\dcp$, 
if $\sin^2\tz\leq0.43~\textrm{or}~\geq0.58$. A $3\sigma$ discovery is possible if 
$\sin^2\tz\leq0.39~\textrm{or}~\geq0.62$ with the same set of runs.

\subsubsection*{Acknowledgments}
SKA would like to thank Pilar Hern\'andez and Mariam Tortola for useful discussions.
SKA acknowledges the support from the Spanish Ministry for Education and Science projects FPA2007-60323 and FPA2011-29678;
the Consolider-Ingenio CUP (CSD2008-00037) and CPAN (CSC2007-00042); the Generalitat Valenciana (PROMETEO/2009/116);
the European projects LAGUNA (Project Number 212343) and the ITN INVISIBLES (Marie Curie Actions, PITN- GA-2011-289442).

\begin{appendix}

\section{Events vs. $\dcp$}
\label{appendix1}

In this appendix, we consider the variation of appearance
events as a function of $\dcp$ in both $\nu$ and $\anu$ modes.

\begin{figure}[H]

        \begin{tabular}{lr}
                \hspace*{-0.95in} \includegraphics[width=0.8\textwidth]{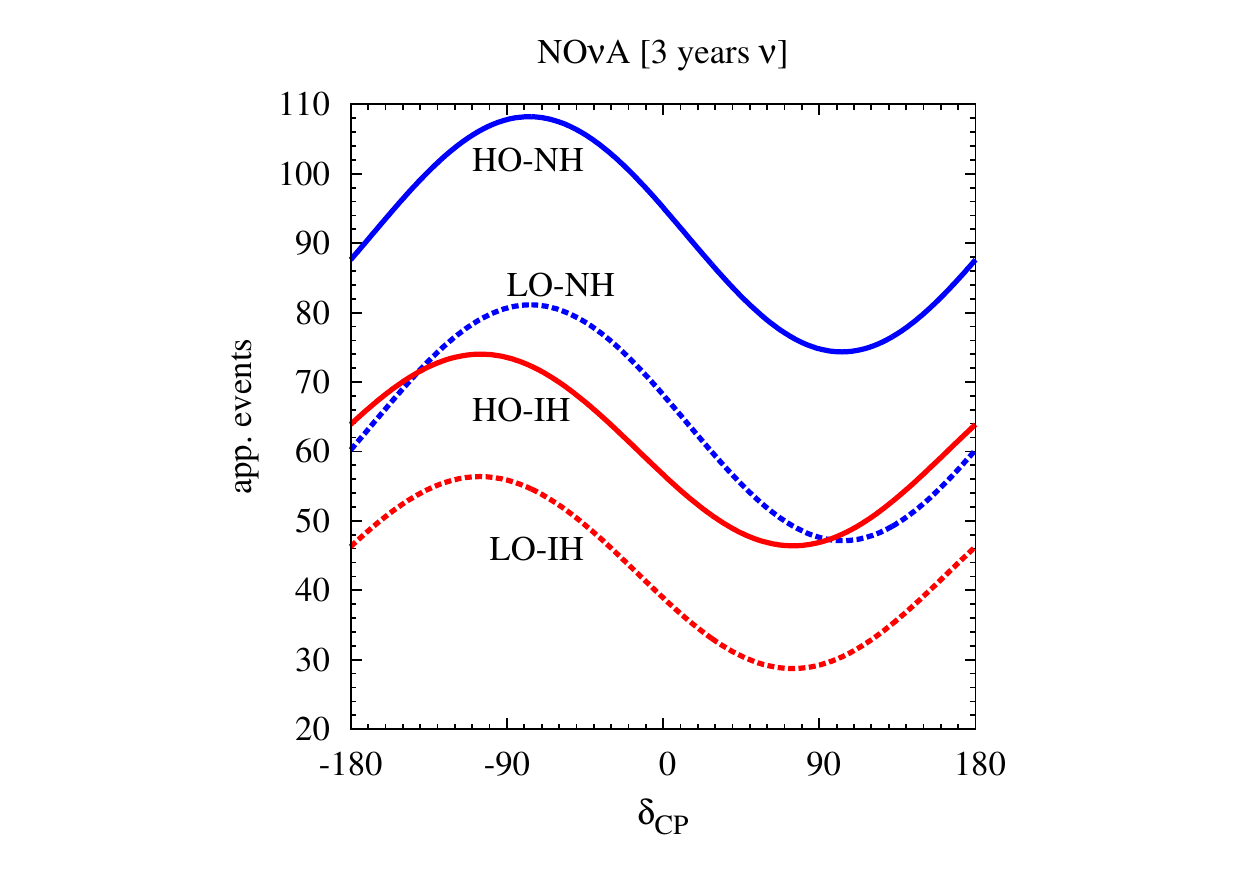}
                & 
                \hspace*{-2.0in} \includegraphics[width=0.8\textwidth]{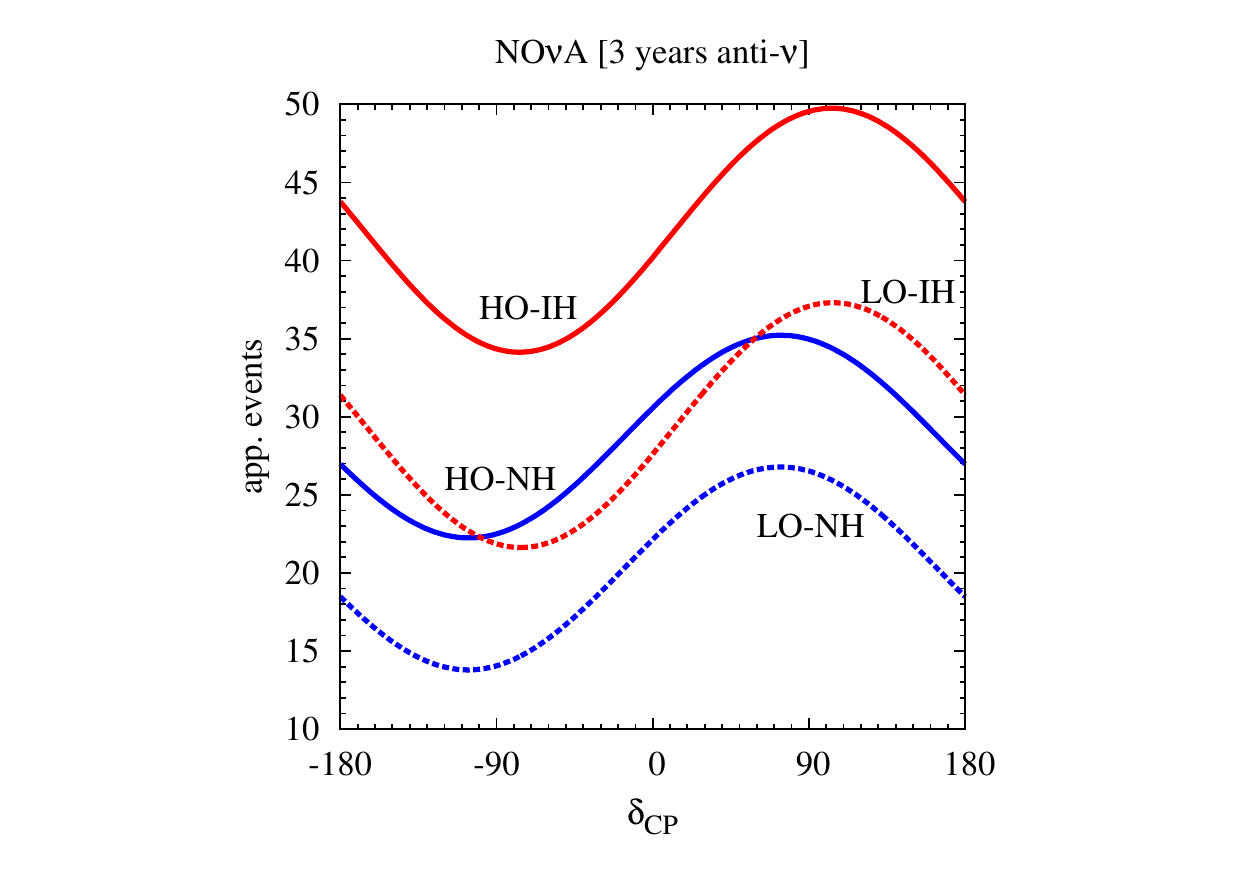}
        \end{tabular}
        
\caption{\label{eventsnova}\footnotesize Total appearance events rates for all possible
combinations of octant and hierarchy as a function of the $\dcp$. The left (right) panel 
is for $\nu$ ($\anu$) running. These plots are for \nova (L=810 km), $\sin^22\ty = 0.089$. 
For LO (HO), $\sin^2\tz = 0.41~(0.59)$.}

\end{figure}

\begin{figure}[H]

        \begin{tabular}{lr}
                \hspace*{-0.95in} \includegraphics[width=0.8\textwidth]{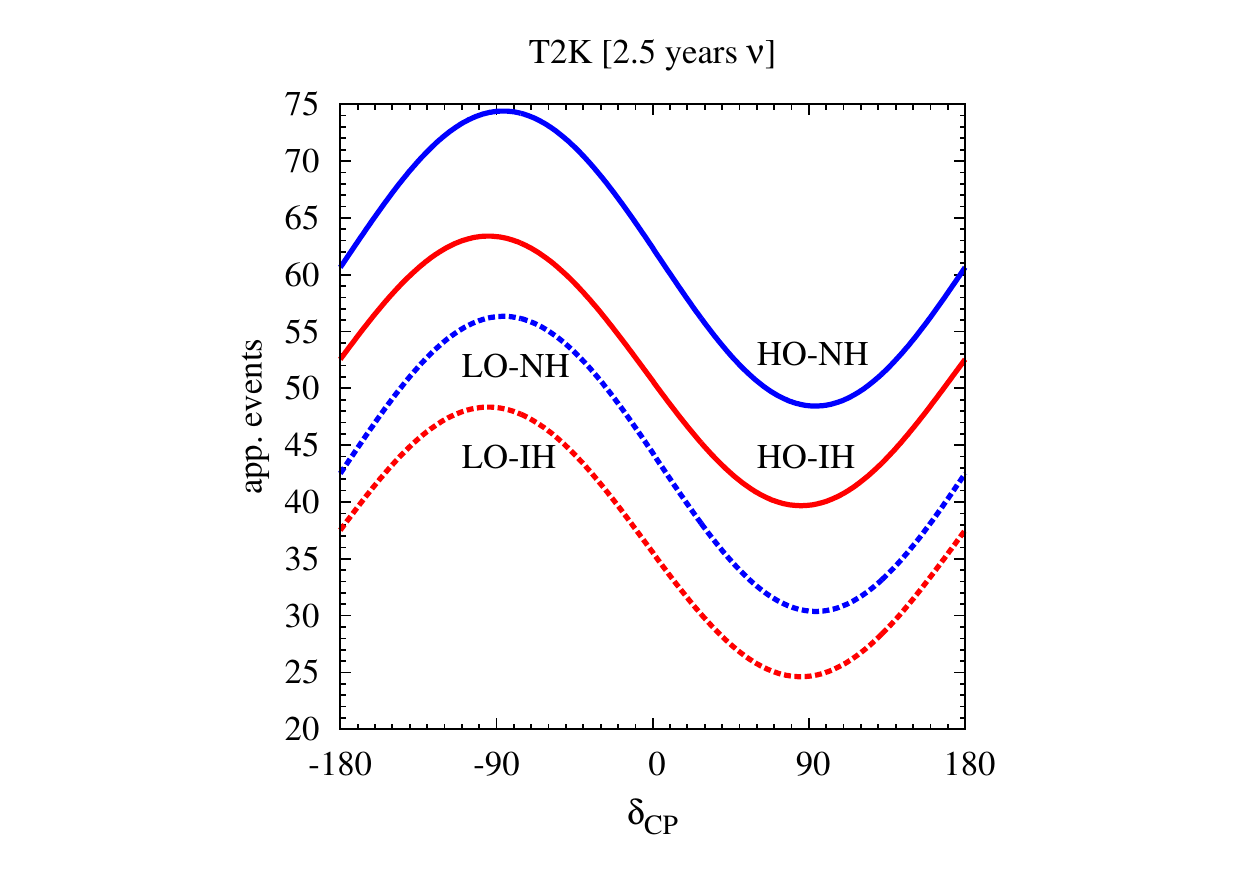}
                & 
                \hspace*{-2.0in} \includegraphics[width=0.8\textwidth]{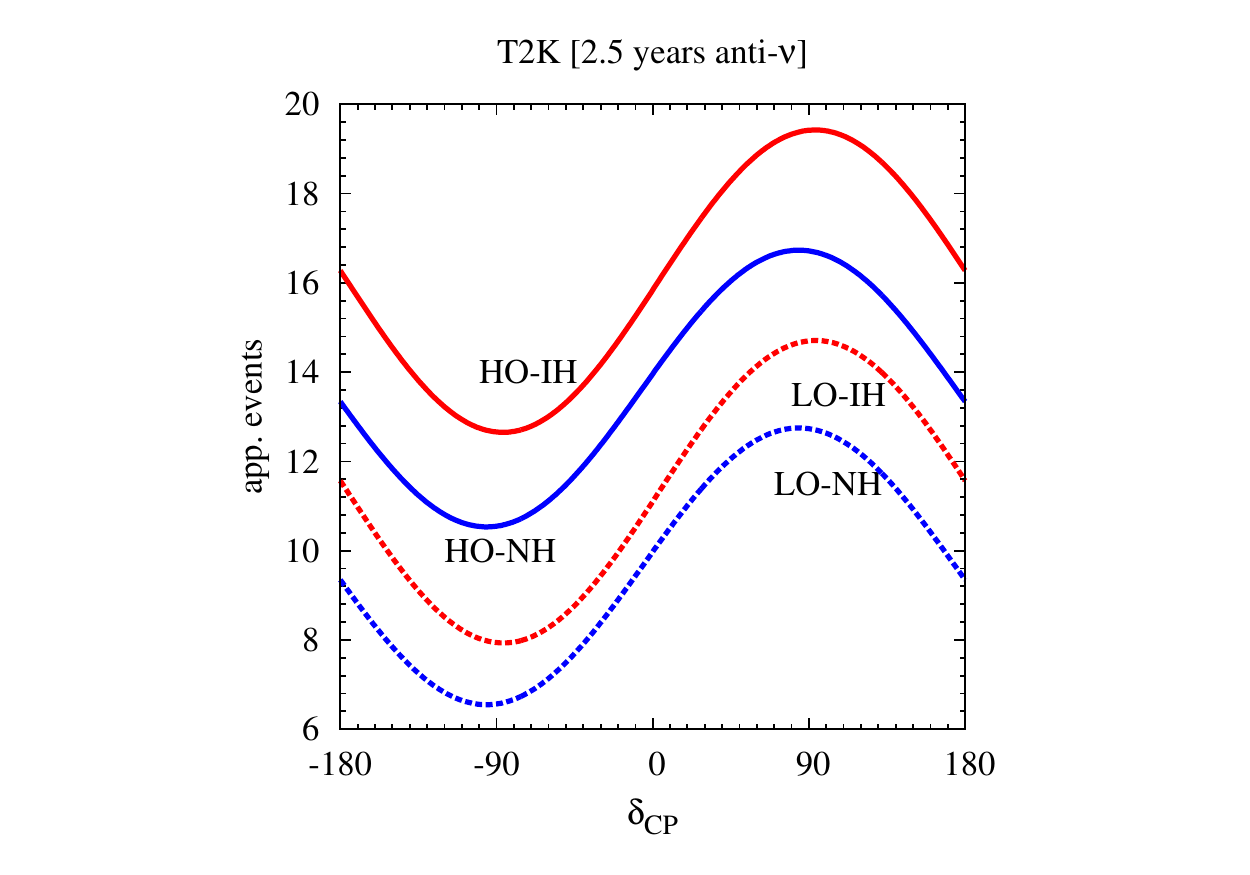}
        \end{tabular}
        
\caption{\label{eventst2k}\footnotesize Total appearance events rates for all possible
combinations of octant and hierarchy as a function of the $\dcp$. The left (right) panel 
is for $\nu$ ($\anu$) running. These plots are for T2K (L=295 km), $\sin^22\ty = 0.089$. 
For LO (HO), $\sin^2\tz = 0.41~(0.59)$.}

\end{figure}

In figure \ref{eventsnova}, we show the variation of $\nu_{\mu}\rightarrow\nu_{e}$ appearance
events vs. $\dcp$ for \nova in both $\nu$ and $\anu$ modes. We see from the left panel ($\nu$ events)
that the combinations HO-NH and LO-IH are well separated but the other two combinations- HO-IH and
LO-NH have essentially the same event numbers. But in the right panel ($\anu$ events), HO-IH and LO-NH
are well separated and the other two combinations are nearly degenerate. Thus, we see that the unfavorable
combinations in $\nu$ mode are favorable in $\anu$ mode and vice-verse. 
This feature is seen in the event rates for T2K also as shown in figure \ref{eventst2k} where 
we have assumed T2K to have 2.5 years of each neutrino and anti-neutrino run .

\section{Allowed regions in test $\dcp$ - test $\sin^2\tz$ plane}
\label{errorplots}

\begin{figure}[H]

        \begin{tabular}{lr}
                \hspace*{-0.95in} \includegraphics[width=0.8\textwidth]{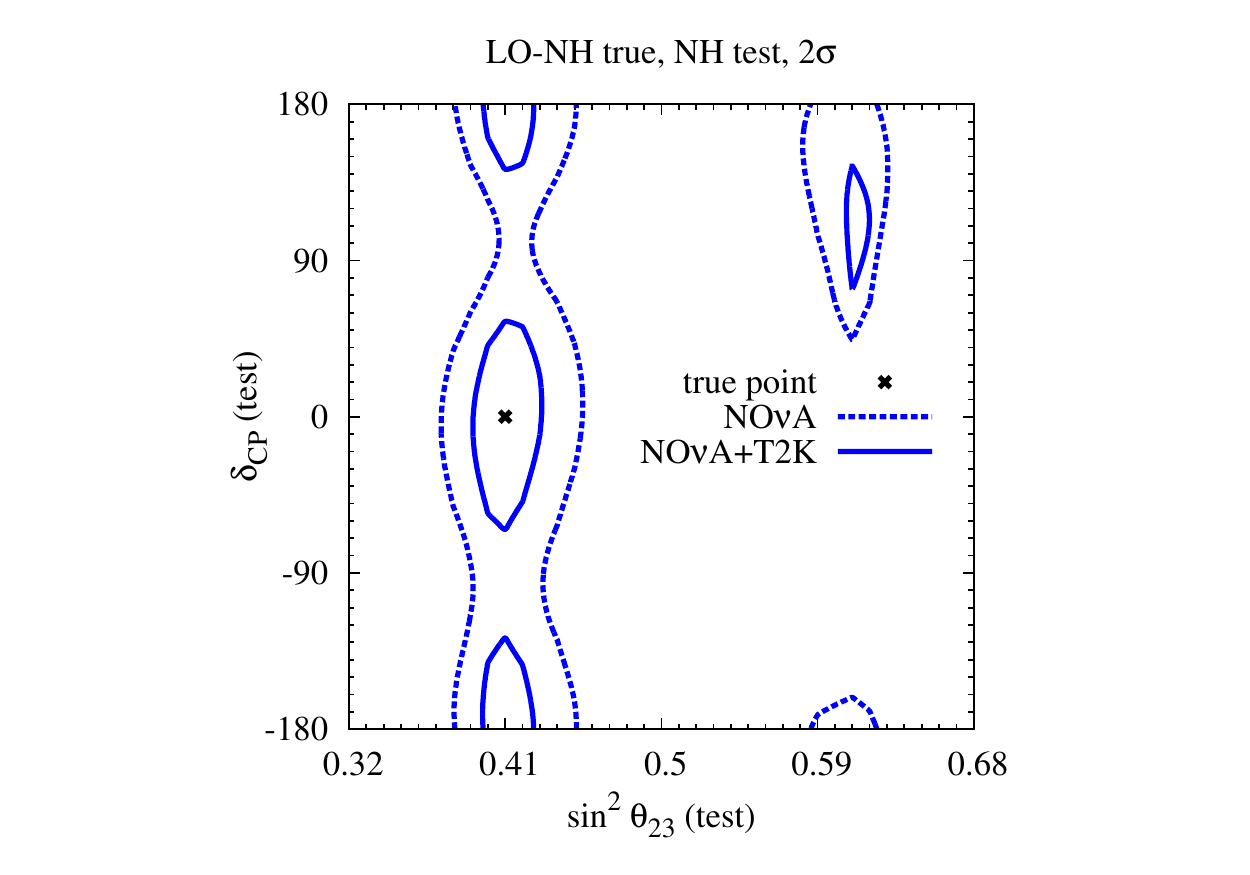}
                & 
                \hspace*{-2.0in} \includegraphics[width=0.8\textwidth]{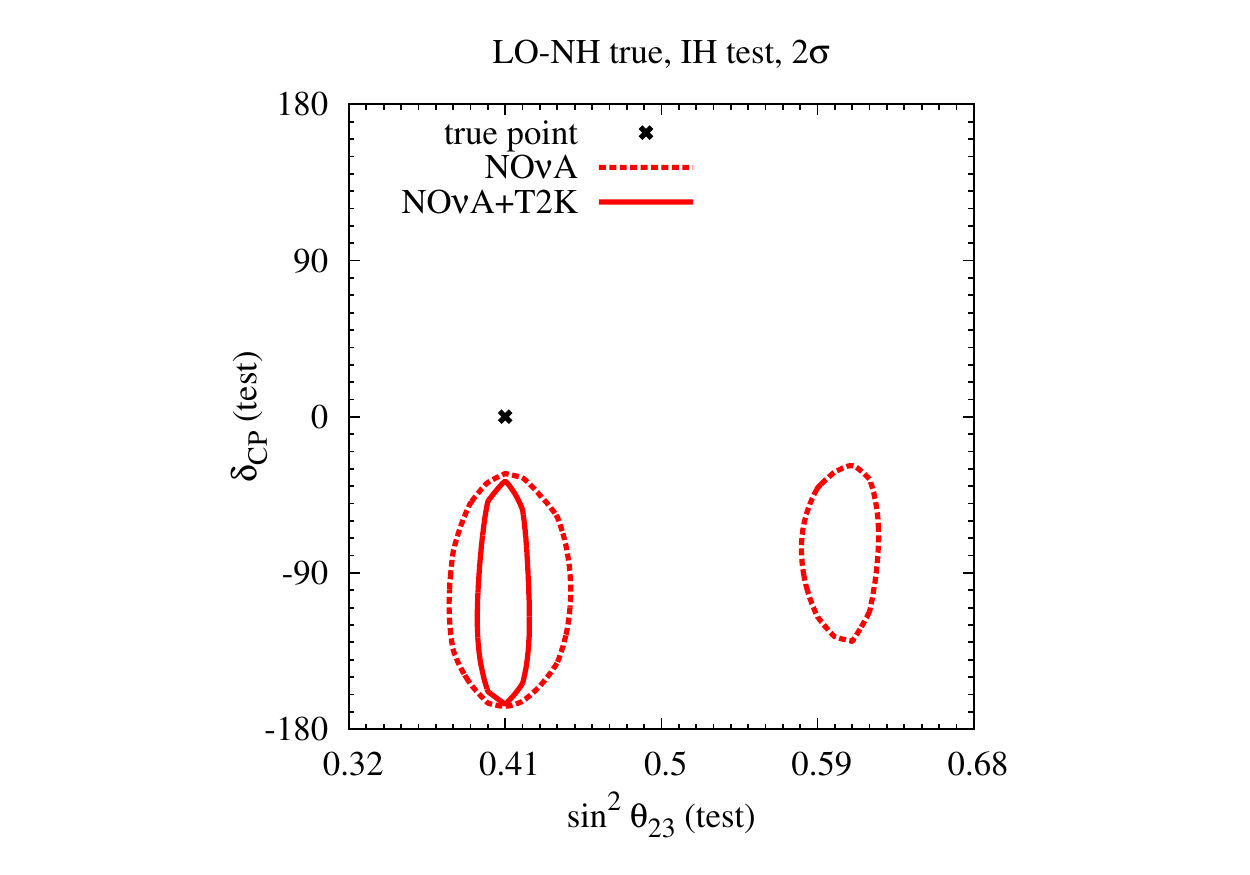}
        \end{tabular}
        
\caption{\label{peanutLONH}\footnotesize Allowed regions in test $\dcp$ - test $\sin^2\tz$ plane 
at $2\sigma$ (2 d.o.f.) C.L. for true $\dcp=0$. LO-NH is assumed to be the true combination. 
The left (right) panel corresponds to NH (IH) being the test hierarchy.}

\end{figure}

\begin{figure}[H]

        \begin{tabular}{lr}
                \hspace*{-0.95in} \includegraphics[width=0.8\textwidth]{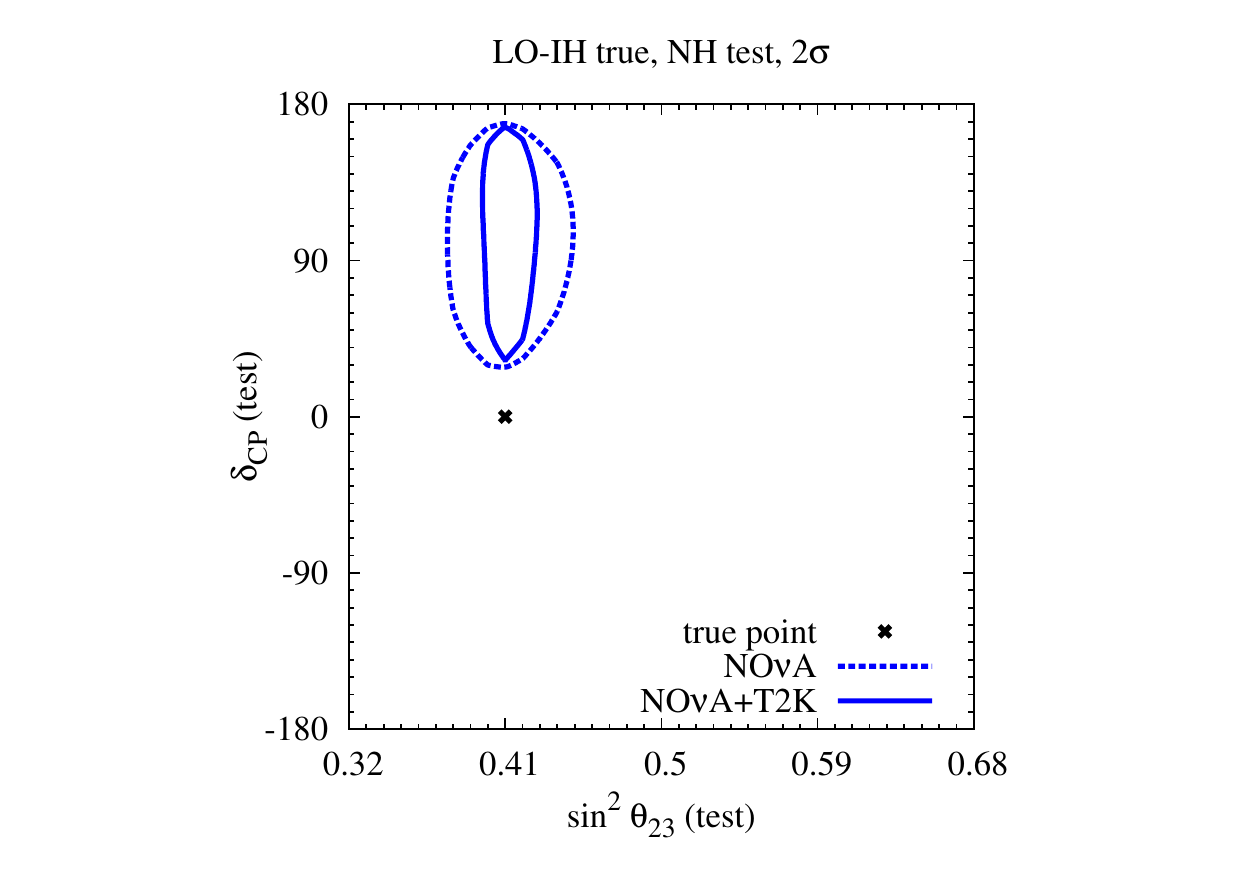}
                & 
                \hspace*{-2.0in} \includegraphics[width=0.8\textwidth]{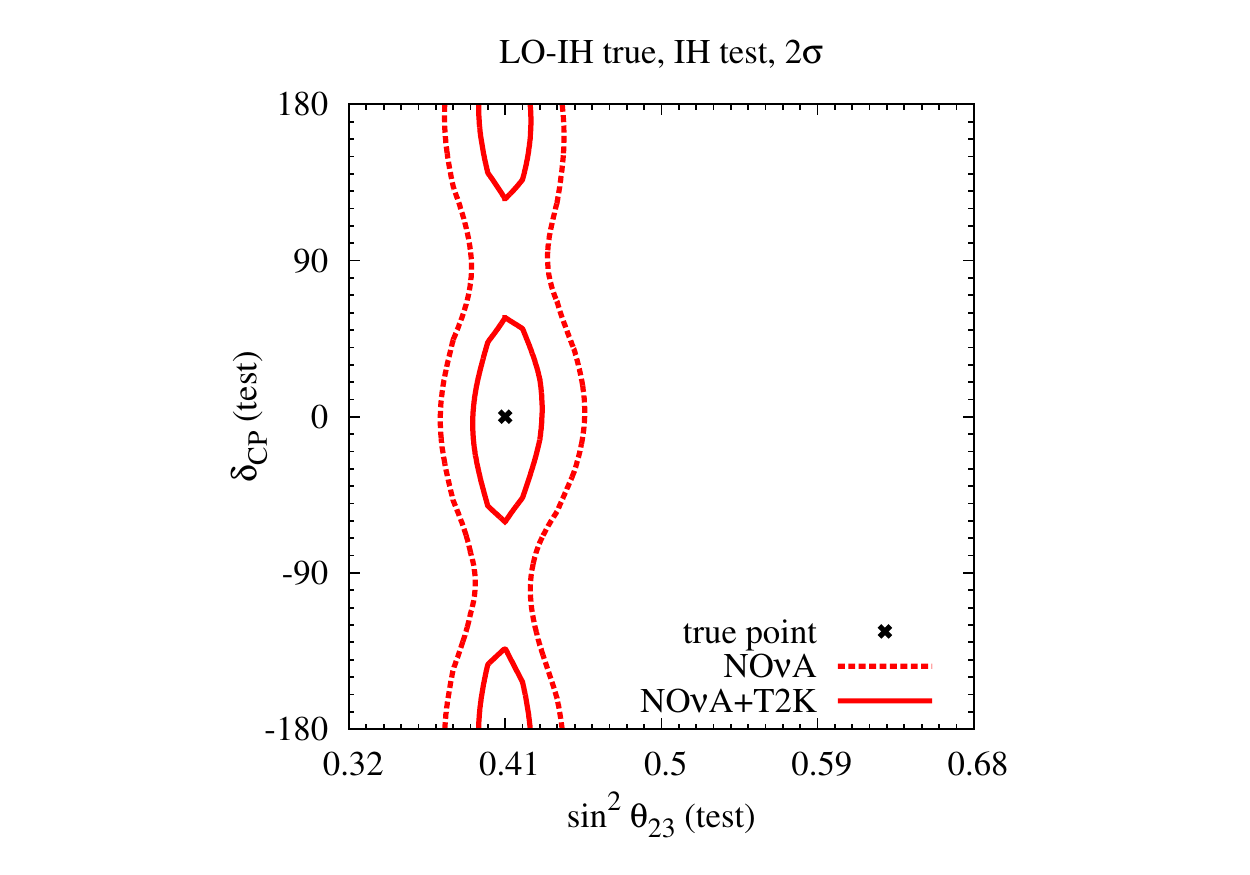}
        \end{tabular}
        
\caption{\label{peanutLOIH}\footnotesize Allowed regions in test $\dcp$ - test $\sin^2\tz$ plane 
at $2\sigma$ (2 d.o.f.) C.L. for true $\dcp=0$. LO-IH is assumed to be the true combination. 
The left (right) panel corresponds to NH (IH) being the test hierarchy.}

\end{figure}

\begin{figure}[H]

        \begin{tabular}{lr}
                \hspace*{-0.95in} \includegraphics[width=0.8\textwidth]{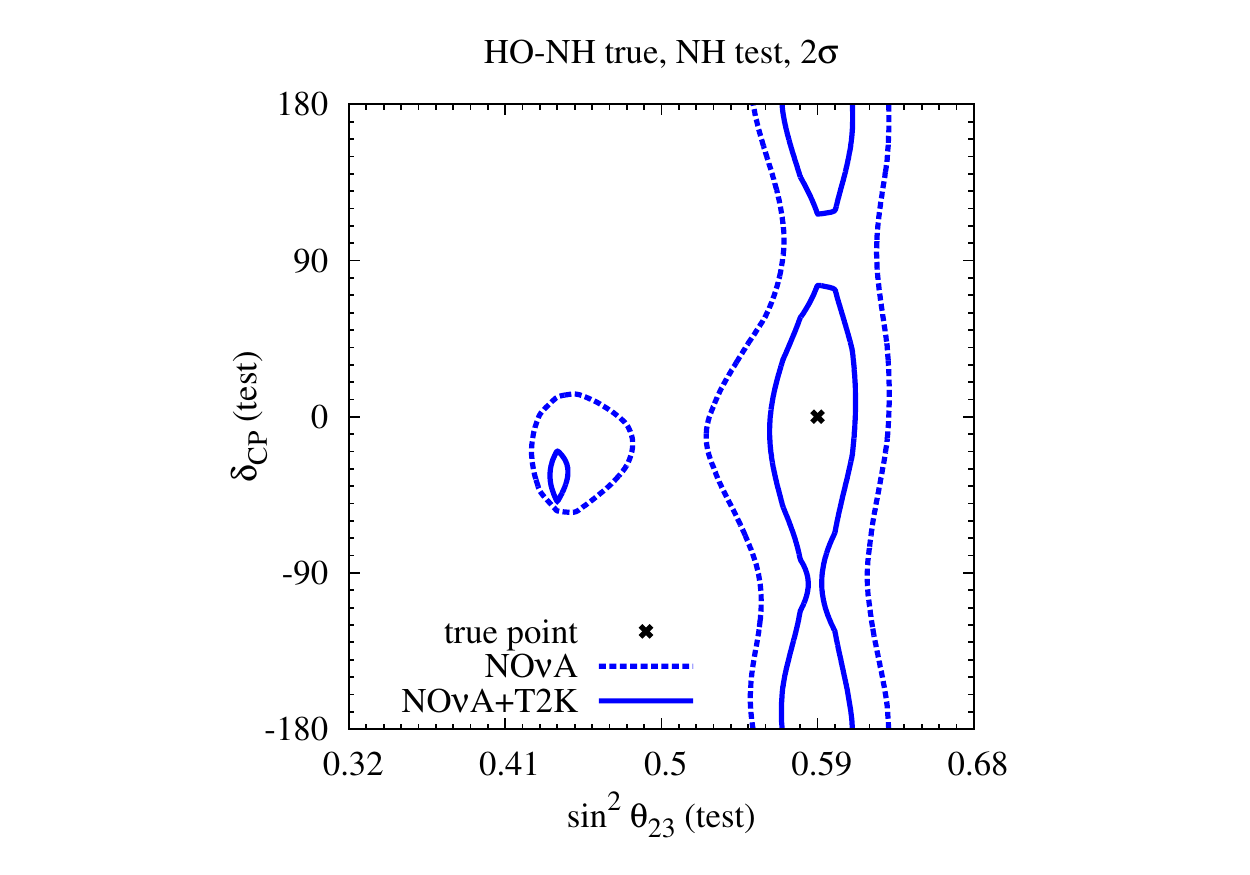}
                & 
                \hspace*{-2.0in} \includegraphics[width=0.8\textwidth]{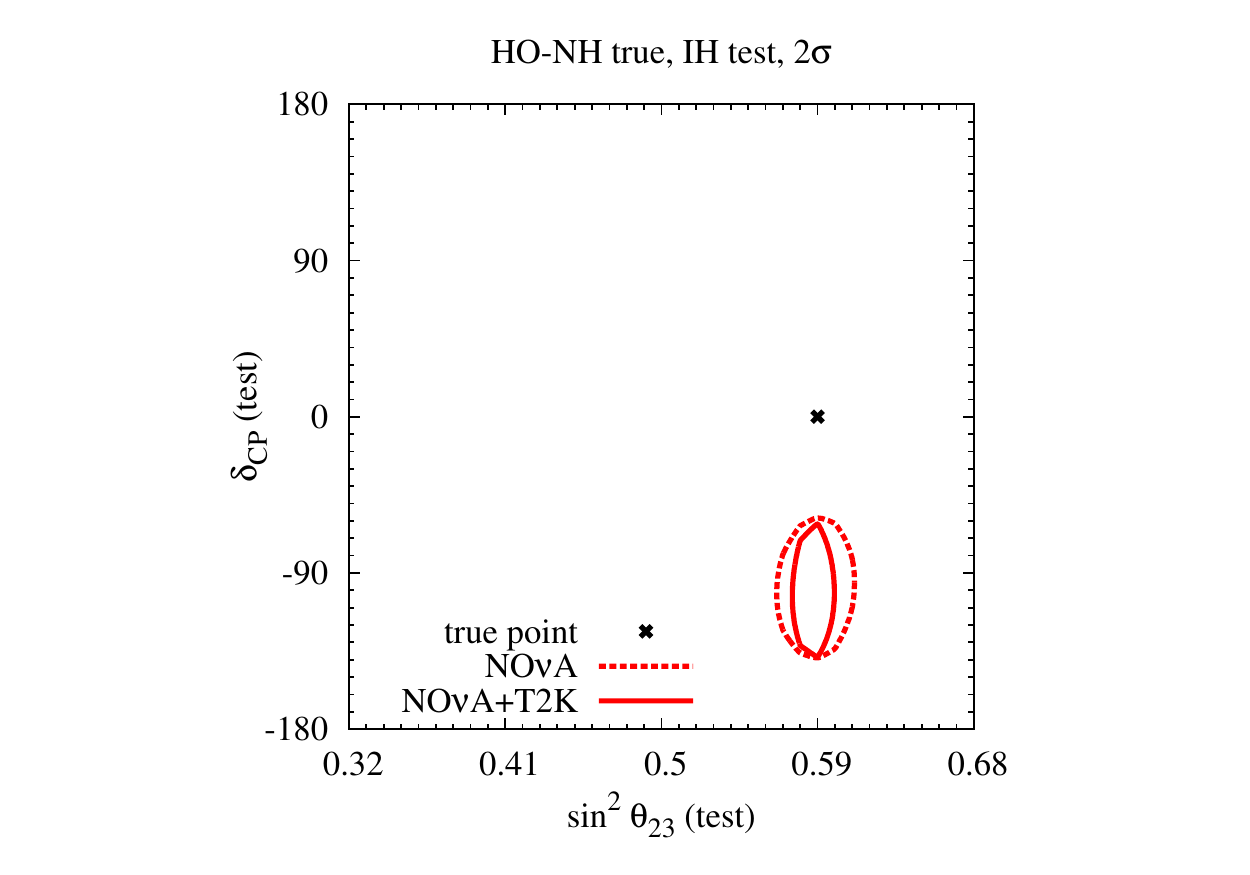}
        \end{tabular}
        
\caption{\label{peanutHONH}\footnotesize Allowed regions in test $\dcp$ - test $\sin^2\tz$ plane 
at $2\sigma$ (2 d.o.f.) C.L. for true $\dcp=0$. HO-NH is assumed to be the true combination. 
The left (right) panel corresponds to NH (IH) being the test hierarchy.}

\end{figure}

\begin{figure}[H]

        \begin{tabular}{lr}
              \hspace*{-0.95in} \includegraphics[width=0.8\textwidth]{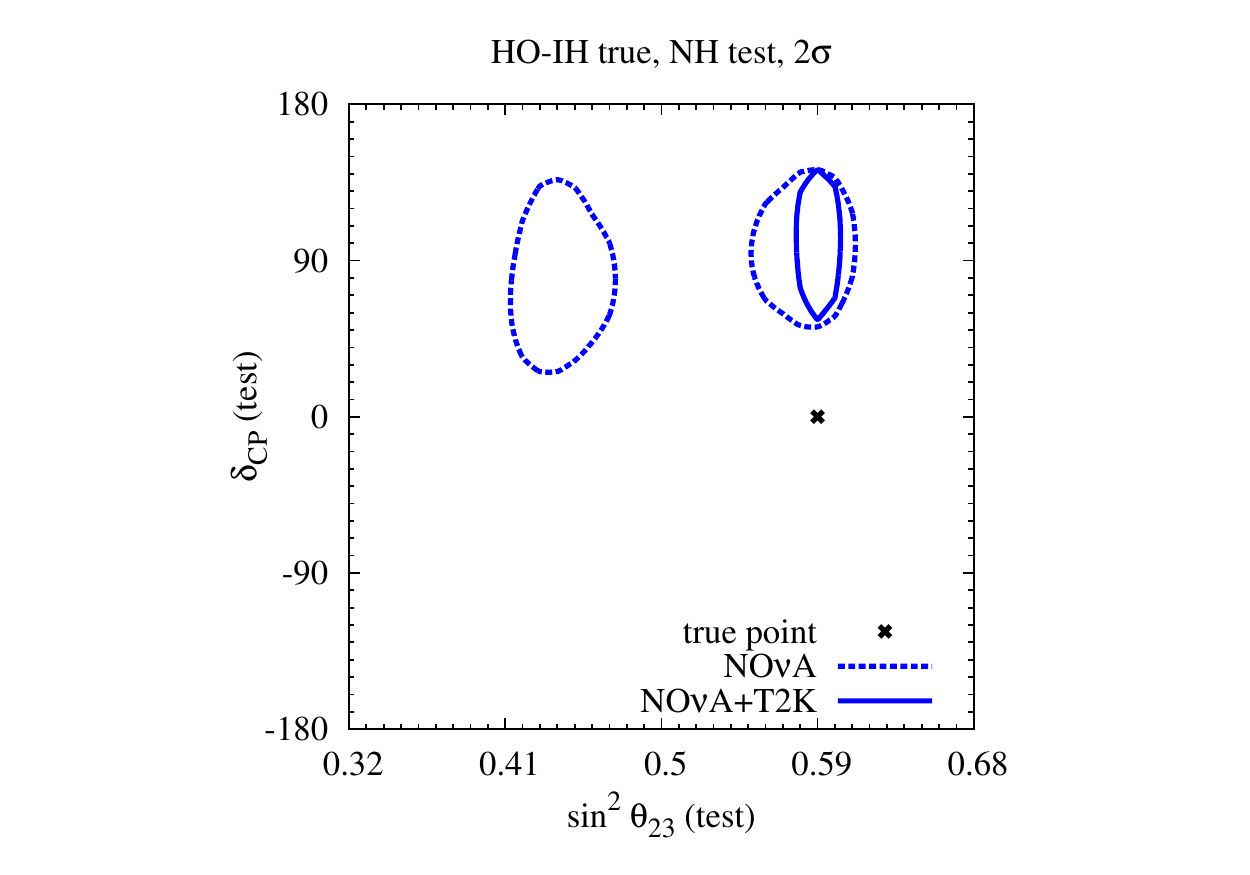}
                & 
              \hspace*{-2.0in} \includegraphics[width=0.8\textwidth]{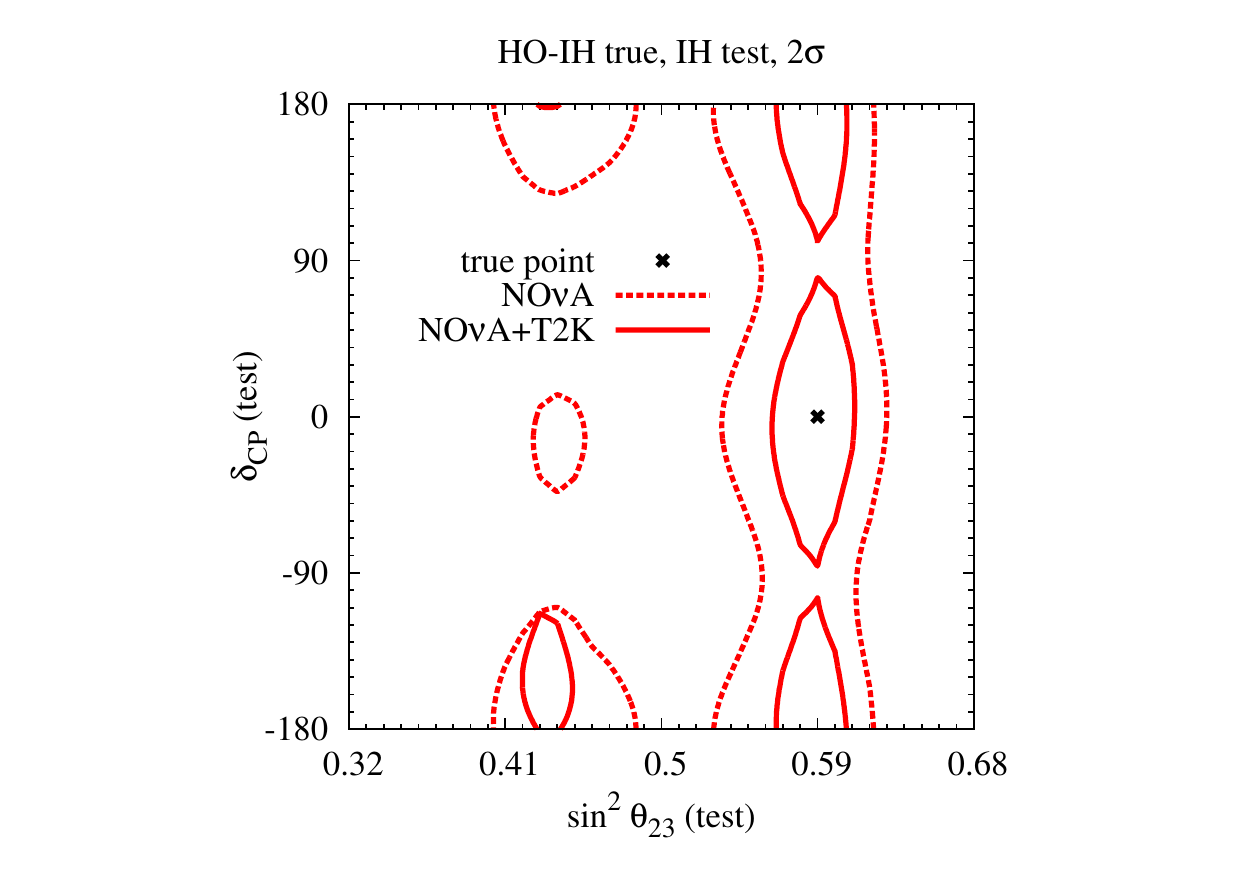}
        \end{tabular}
        
\caption{\label{peanutHOIH}\footnotesize Allowed regions in test $\dcp$ - test $\sin^2\tz$ plane 
at $2\sigma$ (2 d.o.f.) C.L. for true $\dcp=0$. HO-IH is assumed to be the true combination. 
The left (right) panel corresponds to NH (IH) being the test hierarchy.}

\end{figure}

In section \ref{physics}, we saw that the CP conserving values of $\dcp$ pose the biggest challenge
in octant determination.
Therefore, in this appendix, we simulate T2K and \nova data for true $\dcp=0$ and analyze it. 
Figures \ref{peanutLONH}-\ref{peanutHOIH} are drawn for each of the four possible 
combinations of true hierarchy and true octant. These figures show the regions allowed by the data
in the test $\dcp$ - test $\sin^2\tz$ plane at $2\sigma$ ($\Delta \chi^2 \leq 6.18$ for 2 d.o.f.).

These figures illustrate that, in general, the combined data of T2K and \nova discriminate
against the wrong octant of $\tz$. Also, note that, these figures give us the allowed ranges in 
both $\sin^2\tz$ and $\dcp$ (2 d.o.f.) around the right and wrong octant for the single value of true $\dcp=0$.

\end{appendix}

\bibliographystyle{JHEP}
\bibliography{referenceslist}

\end{document}